\documentclass{mn2e}
\usepackage{graphicx}
\pdfoutput=1
\begin{document}

\title[The 2009 radio outburst of XTE J1752$-$223]
      {XTE J1752$-$223 in outburst: a persistent radio jet, dramatic flaring, multiple ejections and linear polarisation}
\author[Brocksopp et al.]
    {C.~Brocksopp$^1$\thanks{email: c.brocksopp@ucl.ac.uk}, 
 S.~Corbel$^2$, A.~Tzioumis$^3$, J.W.~Broderick$^4$, J.~Rodriguez$^2$,  J.~Yang$^5$,\newauthor
R.P.~Fender$^4$, Z.~Paragi$^5$ \\
$^1$Mullard Space Science Laboratory, University College London, Holmbury St. Mary, Dorking, Surrey RH5 6NT, UK\\
$^2$Laboratoire AIM (CEA/IRFU - CNRS/INSU - Universit\'e Paris Diderot), CEA DSM/IRFU/SAp, F-91191 Gif-sur-Yvette, France\\
$^3$Australia Telescope National Facility, CSIRO, PO Box 76, Epping, NSW 1710, Australia\\
$^4$School of Physics and Astronomy, University of Southampton, Southampton, Hants SO17 1BJ, UK\\
$^5$Joint Institute for VLBI in Europe, Postbus 2, 7990 AA Dwingeloo, The Netherlands\\
}
\date{Accepted ??. Received ??}
\pagerange{\pageref{firstpage}--\pageref{lastpage}}
\pubyear{??}
\maketitle
\begin{abstract}
The black hole candidate, XTE~J1752$-$223, was discovered in 2009 October when it entered an outburst. We obtained radio data from the Australia Telescope Compact Array for the duration of the $\sim$9 month event. The lightcurves show that the radio emission from the compact jet persisted for the duration of an extended hard state and through the transition to the intermediate state. The flux then rose rapidly by a factor of 10 and the radio source entered a series of at least 7 maxima, the first of which was likely to be emission associated with the compact jet. The subsequent 6 flares were accompanied by variable behaviour in terms of radio spectrum, degree of linear polarisation, morphology and associated X-ray behaviour. They were, however, remarkably similar in terms of the estimated minimum power required to launch such an ejection event. We compare the timing of radio peaks with the location of the ejecta, imaged by contemporaneous VLBI experiments. We then discuss the mechanism behind the events, in terms of whether discrete ejections is the most likely description of the behaviour. One ejection, at least, appears to be travelling with apparent superluminal motion. The range of properties, however, suggests that mutiple mechanisms may be relevant and that at least some of the emission is coming from shocked interactions amongst the ejecta and between the ejecta and the interstellar medium. We also compare the radio flux density with the X-ray source during the hard state and conclude that XTE~J1752$-$223 is a radio-weak/X-ray-bright outlier on the universal correlation for black hole transient sources.

\end{abstract}

\begin{keywords}
stars: individual: XTE~J1752$-$223 --- accretion, accretion discs --- X-rays: binaries --- radio continuum: stars
\end{keywords}
\section{Introduction}
Astrophysical jets are now thought to be standard features of black hole X-ray transient events; they are no longer assumed to be associated with ``unusual'' systems. They are launched from the central regions of the accretion disc, which surrounds the black hole, and take different forms. The jet may be a partially self-absorbed, approximately conical outflow which transports a large percentage of the accretion power away from the system. At other times it may be in the form of very bright, discrete ejecta which can sometimes be imaged as they travel away from the binary system. The relationship between the jets and accretion state has been a subject of intense study for a number of years but it is not yet clear exactly what conditions are required to launch either type of jet.

In a canonical outburst there are compact radio jets present, displaying a flat or inverted spectrum ($\alpha \ge 0$, where the flux density, $S_{\nu}\propto\nu^{\alpha}$ and $\nu$ is the frequency of observation), while the X-ray source is in its hard spectral state; this state is typified by a hard power-law X-ray spectrum and a high level of aperiodic X-ray variability. The X-ray emission is thought to be dominated by some sort of corona, or base of a jet close to the black hole. As the outburst evolves, the X-ray spectrum becomes increasingly dominated by thermal emission from the accretion disc and the level of aperiodic variability drops. The compact jet persists through this intermediate state, in some cases giving way to an optically thin major radio flare at the point of spectral transition. The jet is then expected to be quenched, perhaps as a result of some sort of equatorial disc wind (Neilsen \& Lee 2009; Ponti et al. 2012), as the X-ray source enters its soft state. Finally, as the X-ray outburst decays and returns to the hard state, the steady radio jet again returns with a flat spectrum and relatively low flux density at radio wavelengths (Corbel et al. 2004; Fender, Belloni, Gallo 2004; Fender 2006; Belloni 2010; and references therein). 

As well as associations between the jet characteristics and X-ray spectral behaviour, there have also been attempts to link the ejecta with the X-ray timing properties; the ejections typically seem to have a loose -- but not exact -- correspondence with episodes of very low integrated noise variability (Fender, Homan \& Belloni 2009). However, the 2009 outburst of H\,1743$-$322, showed a more definite association between ejecta and timing events; the ejections took place as the X-ray variability began to decrease and the Type C quasi-periodic oscillations disappeared from the power density spectrum (Miller-Jones et al. 2012).

In some sources, however, radio emission is present for the duration of the X-ray outburst. A few sources, such as XTE J1859+226, enter a series of optically thin flaring events, causing the radio flux density to remain high throughout the soft state (Brocksopp et al. 2002). This behaviour is described by Fender, Belloni \& Gallo (2004) as repeated ``crossing the jet line'', as the X-ray source switches between the hard and soft intermediate states and the jet correspondingly enters a series of discrete ejection events. Conversely, XTE J1748$-$288 decayed conventionally but then re-brightened and entered an episode of flaring following the return to the hard state (Brocksopp et al. 2007). The flaring may be because the radio source enters a series of ejection events, or because of interactions between the jet and gas in the surrounding environment (Corbel et al. 2002) or because of collisions between the earlier, slower ejected material and subsequent faster ejecta (Fender, Belloni, Gallo 2004). Alternatively, episodes of rapid flaring in GRS~1915+105 have also been closely related to X-ray behaviour and interpreted as ejection of the X-ray corona (Rodriguez et al. 2008a, b).

Distinguishing between these different sources of optically thin radio emission is not straightforward. The radio emission detected during standard monitoring  is typically spatially unresolved and so cannot be linked to specific imaged ejecta or locations. Some of the radio variability has an analogue in the X-ray lightcurves, some of it in the variability of the X-ray spectrum and some seems to be unrelated to any X-ray activity. Estimates of the power used to launch an ejection may give further insight as to the mechanism. Some of the radio events are associated with specific polarisation properties; in theory this should also aid identification of the flaring mechanism but, in practice, the observation of polarised emission can result in more questions than answers (e.g. Fender 2003 and Brocksopp et al. 2007).

A relatively small number of black hole X-ray transients have been observed with a detectable level of linear polarisation (see Fender 2003 and Brocksopp et al. 2007 for details of the sources). While optically thin synchrotron sources can theoretically be linearly polarised up to 70\% and optically thick sources up to 15\% (Longair 1994), this is rarely seen in practice. The ``missing'' polarisation can be lost via a number of mechanisms, any of which may be applicable in X-ray transient systems. Some unresolved sources are comprised of multiple components such as weakly polarised core and highly polarised jet ejecta (e.g. GRS~1915+105 (Fender et al. 1999), GX~339$-$4 (Corbel et al. 2000) and various radio galaxies (e.g. Cawthorne et al. 1993)). Alternatively different ``packets'' within the source may be polarised equally but with different values of polarisation angle, such as the precessing jets of SS433, resulting in a net reduction in observed polarisation (e.g. Stirling et al. 2004). Finally, the degree of polarisation can also be reduced through rotation of the polarisation angle; this can be independently of the wavelength ($\lambda$) via shocks, rotation of the magnetic field structure or acceleration/deceleration of the jet (see e.g. Fender et al. 2002; Blandford \& K\"onigl 1979) or by being viewed through additional magnetic media causing Faraday rotation, where the degree of rotation is dependent on $\lambda^2$. More recently, Cygnus X-1 has been shown to have linearly polarised X-ray emission, possibly indicating extension of the compact jet out to 400 keV (Laurent et al. 2011).

There is a strong relationship between the X-ray and radio emission of accreting black hole sources during the hard state. A non-linear correlation has been found between the flux densities, $F_{Rad}\propto F_{X}^b$ where $b\sim$ 0.5--0.7 (Hannikainen et al. 1998; Corbel et al. 2003; Gallo, Fender \& Pooley 2003), and this holds both for individual sources and universally across many sources. However, there is now a growing number of ``outliers'' (Coriat et al. 2011; Corbel et al. 2013a), challenging the universality of the correlation and creating a second population of sources (Gallo, Miller \& Fender 2012), still with correlated radio/X-ray emision. There is suggestion that XTE~J1752$-$223 is one of these outliers (Ratti et al. 2012) and we address this further in the paper.

For the remainder of this section we summarise the observations and findings to date, regarding the X-ray transient XTE~J1752$-$223. In Section 2 we describe our observations and in Sections 3--7 we present our results, looking at the lightcurves, radio/X-ray correlation, radio spectrum, linear polarisation and energetics. Finally we discuss these results and their implications in Section 8 and draw our conclusions in Section 9.

\subsection{XTE J1752$-$223}
XTE~J1752$-$223 was discovered on 2009 October 23 by the All Sky Monitor on-board the Rossi X-ray Timing Explorer (ASM/{\sl RXTE}) (Markwardt et al. 2009a). Continued X-ray monitoring by {\sl RXTE}, {\sl MAXI} and {\sl Swift} suggested that the source was a black hole candidate (BHC) with variability indicative of an imminent state transition (Nakahira et al. 2009; Markwardt et al. 2009b; Remillard et al. 2009; Shaposhnikov et al. 2009, 2010). Further analysis of X-ray modelling confirmed the likelihood of the object being a BHC (Nakahira et al. 2012; Reis et al. 2011; Mu{\~n}oz Darias 2010a). 


A bright optical counterpart was found three days after the X-ray discovery, and a bright near-infrared counterpart found almost immediately afterwards (Torres et al. 2009a, b). A radio counterpart was discovered using the Australia Telescope Compact Array (ATCA) with a flux density of $\sim 2$ mJy at both 5.5 and 9 GHz, the flat spectrum consistent with that of the compact jet which is routinely associated with the hard X-ray spectral state (Brocksopp et al. 2009).

On-going X-ray monitoring was complicated for the following two months due to Sun constraints. It appeared that the X-ray source remained in the hard state for an unusually extended period of over two months. Homan (2010) reported the transition to a softer spectral state in mid-January 2010. This was confirmed by the resumption of pointed observations by {\sl RXTE}/PCA and {\sl Swift} which indicated an intermediate state (Shaposhnikov 2010a and Curran et al. 2010 respectively. See also Shaposhnikov et al. 2010b and Nakahira et al. 2010). Follow-up radio observations suggested the on-set of a radio ejection event, often present during spectral state changes (Brocksopp et al. 2010a). Additional radio observations were obtained using the European VLBI Network (EVN) and showed the presence of resolved jet components (Brocksopp et al. 2010b,c).

\begin{figure*}
\hspace*{-2cm}\includegraphics[width=22cm,angle=0]{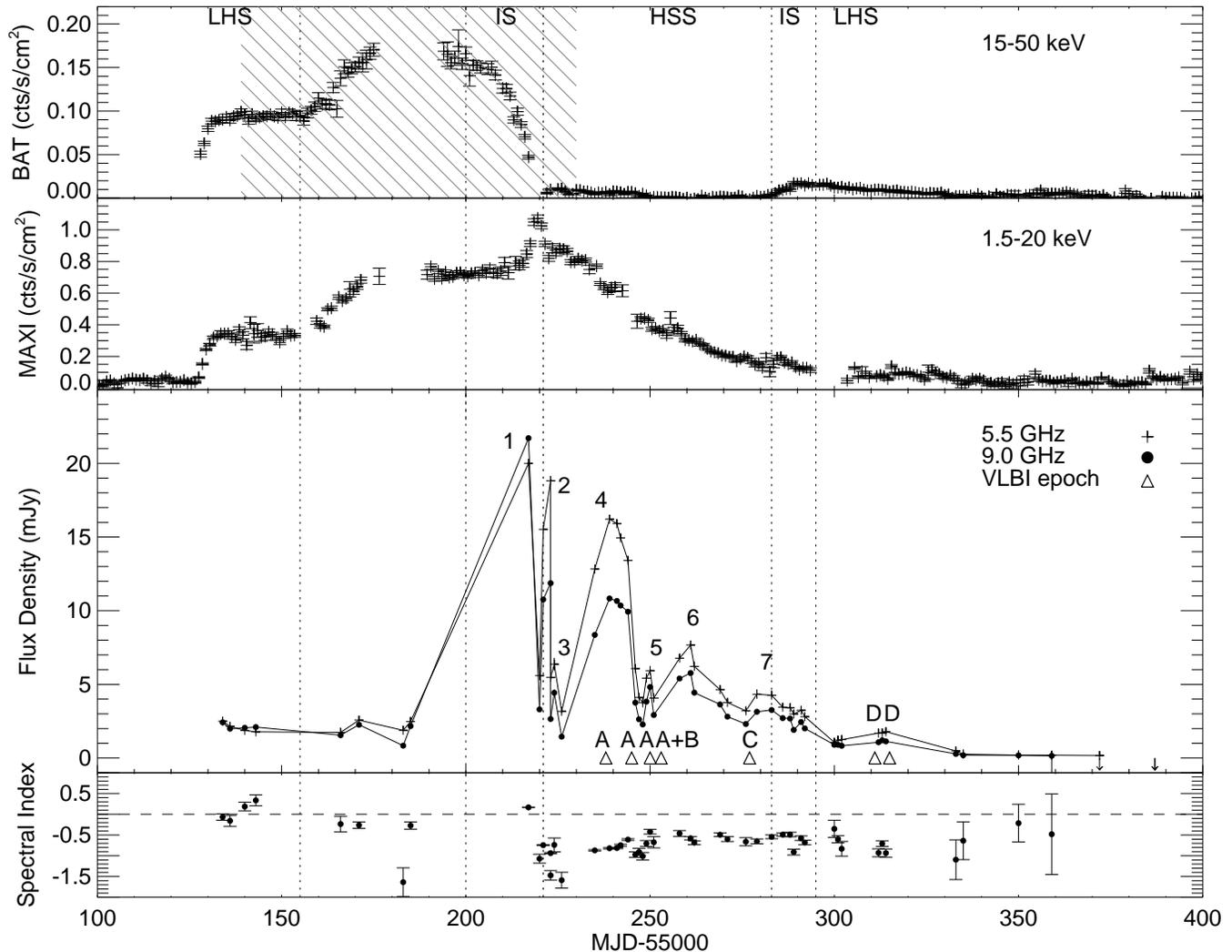}
\vspace*{-2cm}
\caption{X-ray and radio lightcurves for the 2009-2010 outburst of XTE~J1752$-$223. The top panel shows the 15--50 keV BAT data and the second panel shows the 1.5--20 keV data from MAXI. Vertical dotted lines indicate the times of X-ray state change, as defined by Shaposhnikov et al. (2010), although we note that the time of the hard-intermediate state transition was not well-determined and took place some time during the period of sun constraint (hatched region). The third panel shows the ATCA radio data at 5.5 and 9 GHz. Triangles mark the dates of the VLBI observations and labelled A, A+B, C or D according to the ejecta components which were detected (Yang et al. 2011). We add connecting lines between the radio points for clarity but note that caution is needed in use of these lines since there were gaps in our monitoring and we may have missed additional flares. The radio maxima are labelled 1--7 as discussed in the text. Finally, the bottom panel shows the spectral index, $\alpha$, where the flux density, $S_{\nu}\propto\nu^{\alpha}$ and $\nu$ is the frequency of observation.}
\label{lightcurves}
\end{figure*}

The source continued to evolve, returning to the hard state again in 2010 March (Mu{\~n}oz Darias et al. 2010b) and later showing a re-brightening of the infrared counterpart (Buxton et al. 2010). By 2010 August the optical counterpart had faded to a quiescent level (Russell et al. 2010), although re-brightened briefly (Coral-Santana et al. 2010a,b,c).

The radio source was imaged with the European VLBI Network (EVN) and Very Long Baseline Array (VLBA) and a series of ejecta discovered (Yang et al. 2010, 2011). It appeared that there were at least three discrete ejection events (labelled A, B and C) as well as the compact core (labelled D). An accurate identification and position for the radio core were obtained by Miller-Jones et al. (2011) via optical spectroscopy and Very Long Baseline Array (VLBA) radio imaging. Since the position was to the south-east of the radio components detected by Yang et al. (2010), it seemed more likely that both observed ejecta were approaching the observer and the receding components were below the threshold for detection. Further Very Long Baseline Interferometry (VLBI) observations and analysis of the four radio components, and their proper motions, suggested that one component must have experienced significant deceleration; another may provide evidence for XTE J1752$-$223 being a new superluminal source. It also appeared that the radio core was variable (Yang et al. 2011).

Further analysis of the optical and infrared decay showed that there were contributions from both the disc and the synchrotron jet, the latter possibly also associated with an X-ray flare (Russell et al. 2012). Evidence for a synchrotron contribution to the rising optical flux was also presented by Curran et al. (2011), the first time such a hysteresis effect had been found in the optical (as opposed to the infrared, e.g. Buxton \& Bailyn 2004; Russell et al. 2007).

Finally, possible evidence for an X-ray jet was discovered in deep {\sl Chandra} observations presented by Ratti et al. (2012). The imaged component was aligned with the radio jet components of Yang et al. (2011) and at an angular distance of $2\arcsec.9$ from the core. If confirmed, such an X-ray jet component could have been ejected directly from the central source or be shocked emission produced in collisions between ejecta or interaction with the interstellar medium.

\section{Observations}
Radio monitoring at the ATCA commenced on 2009 October 9 and continued until 2010 July 10, giving a total of 52 observations. The new Compact Array Broadband Backend (CABB) was used and observations conducted at 5.5 and 9 GHz . The array was changed into a range of configurations over the course of our programme; the majority of high flux and variability occurred during the 6A and 750B windows. Observing times were typically 1--2 hours but ranged from brief snapshots of minimum 15 minutes of usable data to a maximum of 5 hours.

The ATCA data were flagged, calibrated and imaged in the standard way using {\sc miriad}. The flux and polarization calibrator for all epochs was PKS~1934$-$638; the phase calibrator was IVS~B1752$-$225 (PMN~J1755$-$2232) for the first epoch and MRC~B1817$-$254 (PMN~J1820$-$2528) thereafter. Due to bright sources in the sidelobes at 5.5GHz it was necessary to exclude data from shorter baselines for some of the lower resolution configurations. Thirteen of our observations were brief snapshots for which, following successful deconvolution in the {\sc clean} routine, the {\sc restor} routine was unable to determine a beamsize for reconvolution. In these cases it was necessary to estimate a beam based on adjacent epochs; this had no impact on determination of the flux density, although it did reduce the accuracy of the estimated source position. Additional observations were obtained from the EVN and VLBA. Details of these data can be found in Yang et al. (2010, 2011). 

Publicly available X-ray data were obtained from {\sl Swift}/BAT and {\sl MAXI}/GSC for comparison with the radio data. We also downloaded, reduced and analysed all data obtained with the Rossi X-ray Timing Explorer ({\sl RXTE}) Proportional Counter Array (PCA) during the 2009--2010 outburst of XTE J1752-223. The RXTE/PCA data were reduced following standard steps  (e.g. Rodriguez et al. 2008a) with the {\tt HEASOFT} v6.12 to obtain light curves, spectra and response matrices. Good Time Intervals were defined following the recommended filter criteria\footnote{e.g. http://heasarc.gsfc.nasa.gov/docs/xte/pca\_news.html\#practices}. With a view to producing the Hardness Intensity Diagram (HID) of the source shown later in Fig.~\ref{hid}, we extracted light curves in the standard three energy ranges usually used for this purpose (2.87--5.71 keV, 5.71--9.81 keV and 2.87--20.20 keV), and corrected them for the background. We then extracted source and background spectra from the top layer of PCA/Proportional Counter Unit No. 2, using the appropriate version of the background model according to the brightness of the source. The resultant spectral products were loaded into {\tt XSPEC} v12.7.1 and fitted. As we wanted to obtain fluxes in well defined bands, we fitted the spectra with very simple models, only ensuring that the reduced $\chi^2$ was less than 1.8 for the fit (and fluxes) to be acceptable.

\section{Results - Lightcurves}
We plot the radio lightcurves and spectral indices in the bottom panels of Fig.~\ref{lightcurves}, with the hard X-ray ({\sl Swift}/BAT) and soft X-ray ({\sl MAXI}/GSC) public data in the upper panels. The X-ray data have been discussed in detail by previous authors (Shaposhnikov et al. 2010; Nakahira 2010, 2012) and are just shown here for comparison with the radio.

Radio monitoring of XTE J1752$-$223 started while the X-ray source was in the hard state (Shaposhnikov et al. 2010). The radio source was first detected with flux density $\sim2$ mJy at both 5.5 and 9 GHz, consistent with the flat spectrum that we would expect for a compact jet in the hard state. XTE J1752$-$223 was unusual in that the initial hard state lasted for over a month (55130 to $\sim$55200; Shaposhnikov et al. 2010), as opposed to the few days typical for most black hole X-ray transient sources (but see also e.g. GX~339$-$4 (Corbel et al. 2013a), SWIFT J1753.5-0127 (Cadolle Bel et al. 2007) and SWIFT J174510.8-262411 (Belloni et al. 2012) for sources with hard states of longer duration), and so we were able to detect the compact jet on multiple occasions before the source evolved further. As the X-ray source started its transition to the intermediate state, some time after MJD~55190, the flat spectrum persisted, with the radio flux becoming marginally more variable (see Fig.~\ref{lightcurves}). Once the intermediate state was reached (MJD~55200), the radio source brightened by an order of magnitude and the spectrum remained optically thick (MJD~55217; Peak 1 in Fig.~\ref{lightcurves}). While there is a gap in coverage prior to this peak, the presence of optically thick emission persisting from the compact jet suggests that there were no missed peaks during this gap.

\begin{figure*}
\hspace*{-1cm}\includegraphics[width=21cm, angle=0]{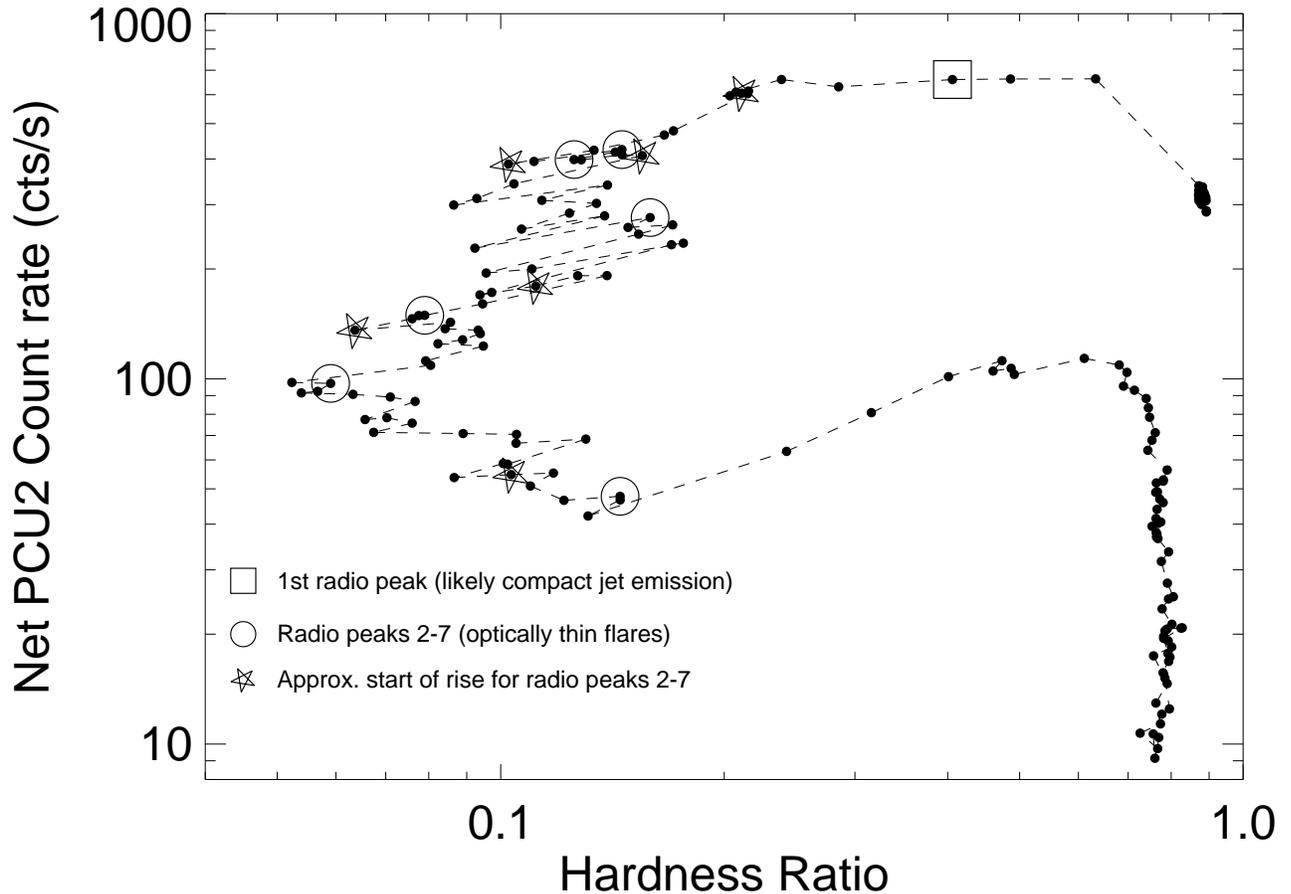}
\vspace{-2cm}
\caption{X-ray hardness:intensity diagram, showing a much higher degree of variability than is obvious in the lightcurves. Symbols have been over-plotted to indicate the times of (i) the first radio maximum at MJD 55217 (square), (ii) the peaks of the subsequent radio flares (open circles) and (iii) estimated beginning of the rise for each of the subsequent radio flares (stars). There is a loose tendency for the radio peaks to be associated with episodes of temporary X-ray spectral hardening but it is by no means a clear relationship.}
\label{hid}
\end{figure*}

As the X-ray source of XTE J1752$-$223 made the transition to the soft state, the initial radio peak (peak 1 in Fig.~\ref{lightcurves}) faded. It then re-brightened almost immediately to $\sim 15-20$ mJy (MJD~55223), this time with an optically thin spectrum which persisted through the rise, peak and decay. This second event was consistent with that of the canonical major radio event often observed at the time of the X-ray state transition. The radio source then entered a series of flares, all optically thin and superimposed on an otherwise decaying lightcurve. In total we observed at least seven maxima, although there were gaps in our coverage and so additional events may have been missed. In contrast, the soft X-rays decay smoothly and the hard X-rays are quenched (although see later in Fig.~\ref{hid}).

It is perhaps remarkable that such dramatic radio flaring did not seem to correlate with features in the X-ray lightcurves. The soft X-ray peak takes place just after the initial radio maximum of MJD~55217 (or up to a couple of weeks later if our radio monitoring missed the true radio peak) and similarly it precedes the second radio peak by a few days. The X-ray decay during the soft state is then relatively featureless, decaying smoothly until the transition to the intermediate state (although see also the following section on correlated behaviour between X-ray and radio emission). The 7th radio peak occurs at the end of the X-ray decay, (see also Corbel et al. 2004) and precedes the hard X-ray reflare by about 10 days, although that does not necessarily rule out any association; peak 7 could indeed represent the return of the compact jets. We discuss the X-ray emission further in the following section.

The source passed through the intermediate state again during its decay to the hard state. Once in the hard state, the radio source continued to vary, with a weak re-brightening during MJD55300--55340. Interestingly, this preceded the X-ray/optical reflare ($\sim$ MJD55350--55375) discussed by Russell et al. (2011) but we did not detect anything contemporaneously with their results, possibly due to lack of sensitivity and less frequent monitoring. We note, however, that a similar radio rebrightening with delayed OIR analogue was detected in GX~339$-$4 (Coriat et al. 2009; see also Miller-Jones et al. 2012). Finally, the radio source of XTE J1752$-$233 remained detectable for over two months before decaying below an upper limit of $\sim 0.03$ mJy.

\section{Results - radio spectrum}
We plot the evolution of the spectral index in the bottom panel of Fig.~\ref{lightcurves}. During the hard state, as expected (Fender 2001), the radio spectrum is flat and partially self-absorbed, with some variation to inverted but, perhaps surprisingly, there is also a single point which is particularly optically thin (MJD~55183). During the transition to the intermediate state, there is a shift to optically thin emission before returning to optically thick at the time of the initial peak in the lightcurve (MJD~55217). Once the X-ray source has completed the transition to the soft state, the radio emission is persistently optically thin and this continues for almost the remainder of the observing program, despite the return of the compact jet in the final hard state.

During the flaring episodes the spectrum remains optically thin but with a degree of variability. The source becomes more optically thin during the events around MJD 55220--55225 (peaks 2 and 3), 55245 (end of peak 4) and 55290, adding weight to a suggestion of multiple ejection events. The temporary episodes of optically thick emission that we might expect prior to each repeated ejection event (e.g. A0620$-$00; Kuulkers et al. (1999) or XTE J1720$-$318; Brocksopp et al. (2005)) are less obvious, but this is unsurprising due to the series of rapid and overlapping events.

The radio spectrum continues to be optically thin even once the X-ray source has returned to the hard state. This would normally suggest the on-going presence of optically thin emission in the environment. In the March 22 image of Yang et al. (2011) there is possible ($\sim 0.75$ mJy) residual emission along the path taken by the jet. However, by April 25 and 29 there is only emission coming from the still-variable core and it is still optically thin. Either the optically thin emission is produced by a flaring core or there is further unresolved emission in the ATCA field of view which has been undetected/resolved out by the VLBI. This would not be surprising, given the low signal-to-noise of the last two VLBI observations. Alternatively it is possible that there is some reason why the jet is unable to build up its flux at the higher energies. We also note that the outlier sources of the radio/X-ray correlation may have less inverted radio spectra (Coriat et al. 2011. See also Corbel et al. 2013b). Therefore it is likely that, even if optically thin, this radio emission represents the onset of the compact jet. Finally, the last two radio detections are consistent with having a flat spectrum.

The improved capabilities of ATCA with the new CABB back-end enable us to split the bandwidth at each frequency, from a single bin of width 2GHz to two bins of width 1GHz. This improves the resolution of our radio spectra, albeit with reduced sensitivity, for those epochs with good signal-to-noise. We obtain similar results to those described above, with linear 4-point spectra. However for consistency between all epochs, we continue to use the 2-point spectra in Fig.~\ref{lightcurves}. In particular, we checked the hard state data of (i) MJD~55183, when the spectrum was particularly optically thin -- the 4-point spectra confirm this result and (ii) the decaying hard state to look for evidence of the moment when the jet re-ignited -- there is no obvious change in spectrum which could indicate a jet turn-on.

\begin{figure}
\hspace*{0cm}\includegraphics[width=9cm]{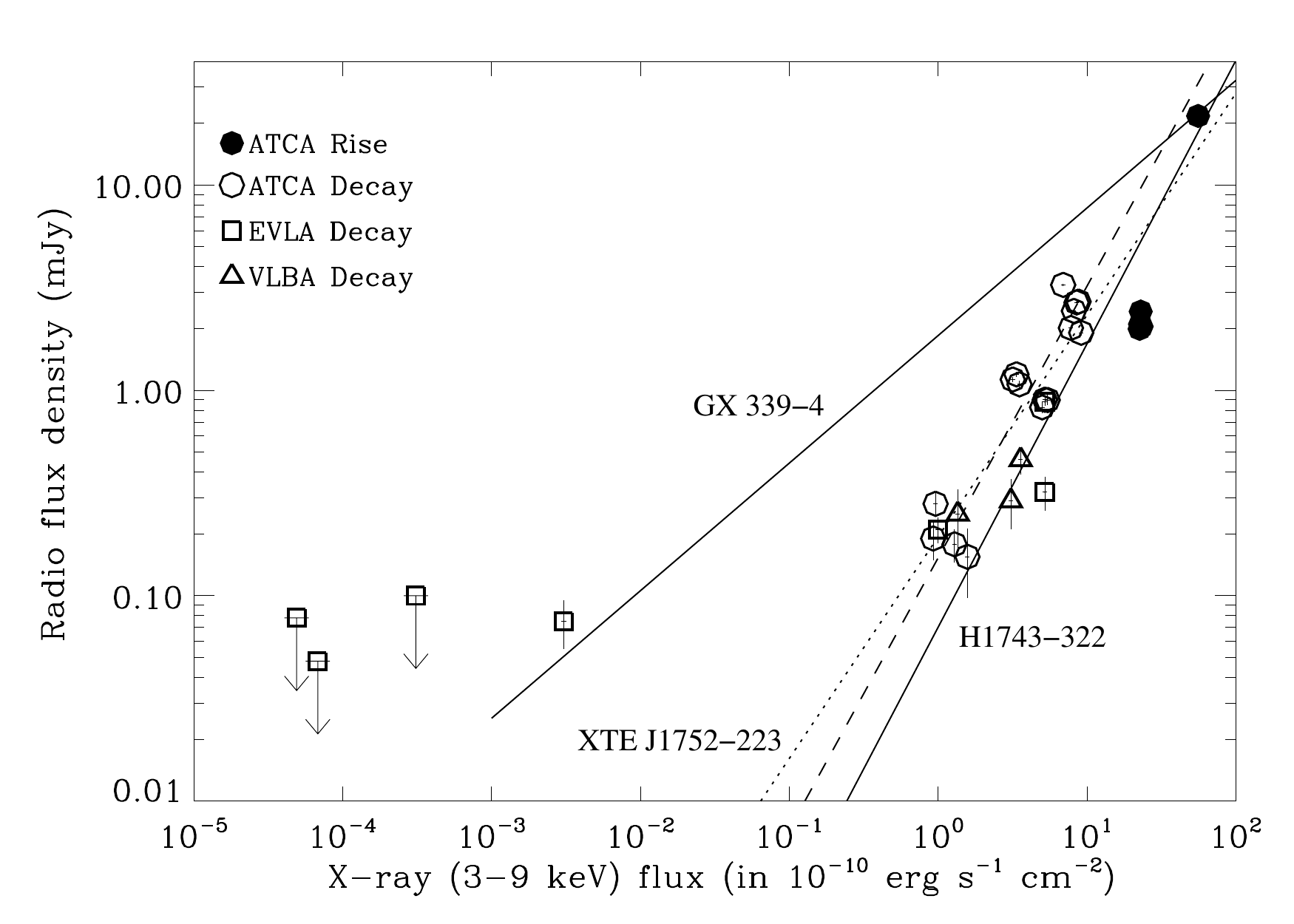}
\vspace{-0cm}
\caption{The universal correlation between X-ray and radio fluxes of black hole transients. XTE J1752$-$223 data from the rising hard state are plotted as filled circles. Data from the decaying hard state are plotted as open circles (ATCA), squares (EVLA; Ratti et al. 2012) and triangles (VLBA; Yang et al. 2011). The standard correlation of GX~339$-$4 (Corbel et al. 2013a) is plotted as the left-hand solid line; the correlation for the outlier, H1743$-$322 (Coriat et al. 2011), is the right-hand solid line. The dotted line is the fit to all XTE J1752$-$223 hard state data, correlation coefficient $1.08\pm0.02$ and the dashed line is the fit to the XTE J1752$-$223 decay data, correlation coefficient $1.31\pm0.04$. This identifies XTE J1752$-$223 as an outlier, joining the main correlation at high and low fluxes only. N.B. the fluxes have not been converted to a common reference distance but, since all three sources are thought to lie at approximately similar distances (Corbel et al. 2013a), they can be considered directly comparable.} 
\label{corr}
\end{figure}

\section{Results - Radio/X-ray correlation}
The lack of features in the X-ray lightcurves compared with the dramatic flaring in the radio lightcurves is perhaps surprising (although not unprecedented, e.g. GX339$-$4; Corbel et al. 2003). This is true of not only the {\sl Swift}/BAT and {\sl MAXI} data presented in Fig.~\ref{lightcurves} but also the pointed X-ray observations presented in e.g. Shaposhnikov et al. (2010); it is not merely an issue of data quality. With this in mind, we present the hardness-intensity diagram in Fig.~\ref{hid}, with the intensity data in the 2.87--20.2 keV energy range and the hardness ratio (5.71--9.5 keV)/(2.87--5.71 keV). A much greater degree of variability is now apparent with a large number of temporary hardening events, not unlike those observed in XTE J1859+226 which were found to be associated with radio events (Brocksopp et al. 2002; Fender et al. 2009).

We have plotted additional symbols on the hardness-intensity diagram in order to indicate significant events in the radio lightcurves. The square corresponds to the first peak in the radio lightcurve, when the X-ray source was still in the intermediate state and the radio emission still produced by the compact jet. The open circles then correspond to the subsequent six radio peaks and the stars correspond to the approximate start of each of these six radio flares. There is no obvious relationship which holds consistent for each flare. The stars do not seem to be associated with either X-ray softenings or hardenings, although obviously the start time of each rise can only be estimated approximately. The radio peaks seem to be very loosely associated with temporary X-ray spectral hardenings but it is far from a clear correlation; not all the circles lie at epochs of spectral hardening and there are many hardening events which apparently are not associated with any change in the radio, although this may be due, in part, to the limited radio coverage. This may suggest that the radio monitoring missed the exact time of the peak or missed other peaks entirely, that the various flares are produced via different mechanisms and/or that the relationship between temporary spectral hardenings and radio flares seen in other sources is not a widespread phenomenon.

Finally we plot the radio and X-ray flux densities for the hard state on the canonical radio flux vs. X-ray flux plot (Fig.~\ref{corr}). The ATCA points are plotted as circles, filled and open for the rise (up to MJD55217) and decay (from MJD55282.5) respectively. EVLA data from the decay are added as open squares and VLBA data as open triangles; we note that the VLBI points are typically fainter than ATCA, most likely on account of the longer baselines resolving out the more extended emission. The dotted line is a fit to all hard state data, giving $L_R \propto L_X^{1.08 \pm 0.02}$. The decay alone gives a fit $L_R \propto L_X^{1.31 \pm 0.04}$ (dashed line). For comparison, the solid lines indicate the correlations for GX~339$-$4 (left-hand line -- standard; Corbel et al. 2013a) and H1743$-$322 (right-hand line -- outlier; Coriat et al. 2011), placing each source at its likely distance of 8 kpc. These results confirm that XTE~J1752$-$223 is indeed an outlier, converging with the standard correlation only at high and low flux. It is possible that the decay was more radio-bright, relative to the X-ray emission, than the rise (see also GX~339-4; Corbel et al. 2013a). We note that the points for the decaying hard state remain optically thin but are likely to represent the compact jet data because i) the X-ray source is in the hard state, ii) the ATCA data are consistent with emission from the resolved core in the VLBI images, iii) the VLBI maps show no resolved components other than the core at this time and iv) Coriat et al. (2011) suggested that the radio spectra of the outlier sources tended to be less flat ($\alpha < 0$) than those on the standard correlation.

\section{Results - linear polarisation}
\begin{figure*}
\includegraphics[width=18cm]{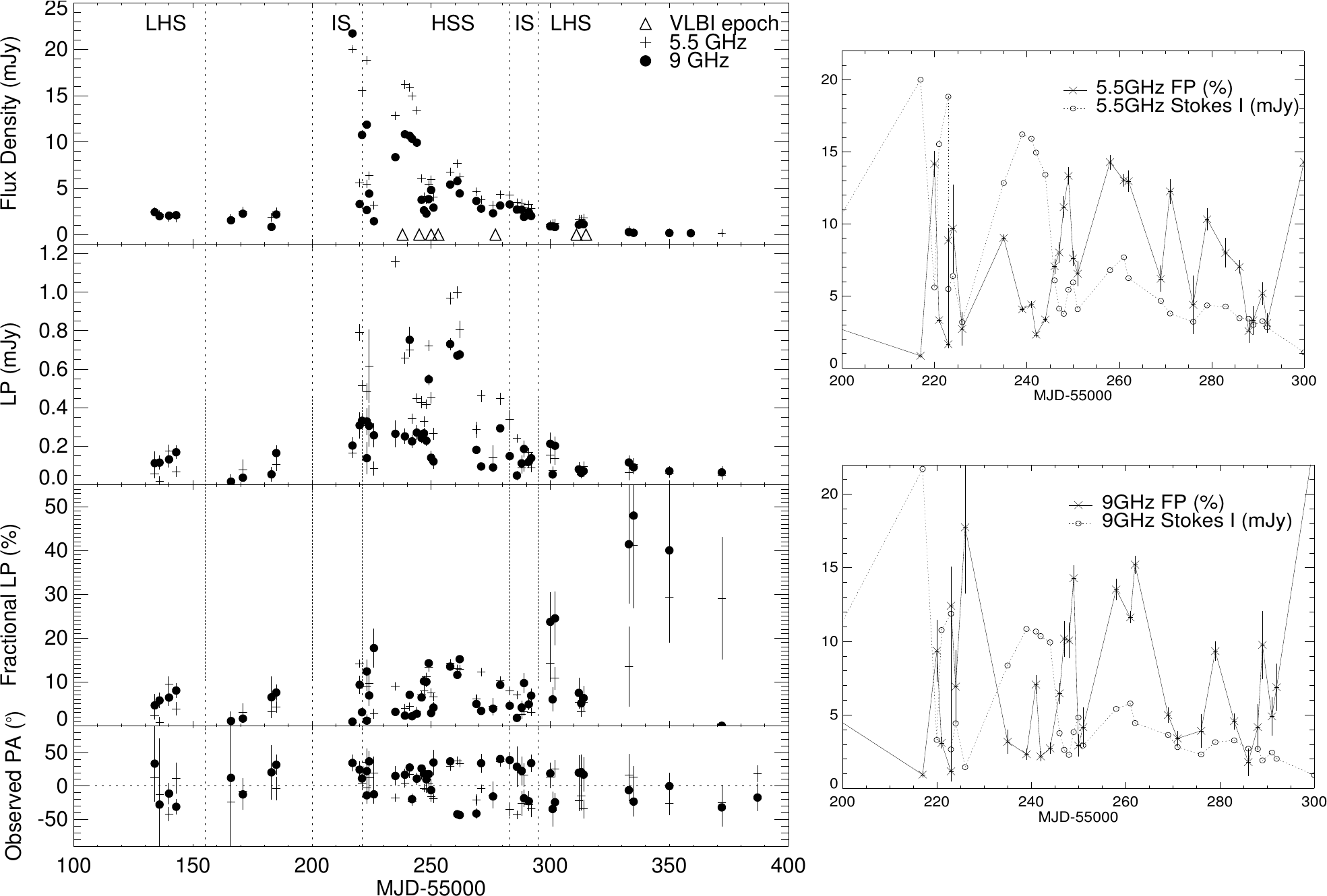}
\caption{LHS: We replot the integrated radio lightcurve in the top panel for comparison. In the three panels below we show the linear polarisation, fractional polarisation (all data included for completeness but the last points unlikely to be reliable) and observed polarisation angle. Crosses indicate the 5.5-GHz data and filled circles the 9-GHz data in each case. RHS: Expanded plots showing the flaring episodes in flux density (crosses) and fractional polarisation (open circles). 5.5-GHz data is shown in the top panel and 9-GHz data in the bottom panel. There is some degree of (anti-)correlation between the various panels but, as discussed in the text, the relationships between the lightcurves is not obvious.}
\label{pol}
\end{figure*}

We obtained images in Stokes $Q$ and $U$ and created linear polarisation maps for all the ATCA epochs, where the linear polarisation, LP, is equal to $\sqrt{Q^2+U^2}$. Most of the data were of the order 10--20 per cent polarised. To illustrate the variability and how it compares with that of the Stokes $I$ lightcurve, we plot the fractional polarisation (FP) in Fig.~\ref{pol}.

During the hard state the FP reaches up to $8.05\pm1.73\%$, relatively high for this state. As the source evolves, the initial peak of the Stokes $I$ is notably weakly polarised, if at all. The FP then increases simultaneously with the decay in flux of the inital peak and transition to the soft state. Once the radio emission becomes optically thin, the (anti-)correlation between Stokes $I$ and FP emission is less pronounced. Peaks 2 and 5 seem to be anti-correlated with the Stokes I flux density, although conversely it could be a delayed correlation. Peaks 3, 6 and 7 seem to have equivalent peaks in the FP. Peak 4 is more complex, showing associated peaks in the FP which are neither correlated or anti-correlated.

We also determined the observed polarisation position angles, using PA$=(1/2)\arctan(U/Q$) and plot them in the bottom panel of Fig.~\ref{pol}. There is a wide range of values -- a very simple, illustrative attempt to quantify the range by fitting a straight line to the data gives a value of $\chi^2_{\nu}>25000$, as we would expect for multi-frequency data and a non-zero rotation measure (RM). We then attempted to calculate the intrinsic PA (PA$_i$) and RM, where PA=PA$_i$+RM$\lambda^2$. This information would have been useful for determining any evolution and/or wavelength-dependency to the linear polarisation. Unfortunately the resultant values of the rotation measure had high scatter and were therefore meaningless, on account of having observations at only two frequencies for the straight-line fits. We were also unable to use the 4-point spectra derived in Section 4 due to insufficient sensitivity. Given the size of the uncertainties on the PA, there is no useful information we can obtain from these data.

\section{Results - Energetics}
There has been only a small number of previous X-ray transient events for which the radio lightcurve during flaring episodes was sufficiently well-sampled to allow estimates of the minimum power required to launch the ejection. These include XTE~J1859+226 (Brocksopp et al. 2002), GRS~1915+105 (Fender et al. 1999), XTE J1748$-$288 (Brocksopp et al. 2007). It is even more unusual to be able to do so for multiple events. Details of these flares can be found in Table~\ref{flares}.

   \begin{table*}
\caption{Summary of the radio events, listing dates of the peaks, maximum observed flux densities and X-ray spectral state at the time of the observed peak. We note that the actual peaks may be different from those observed due to gaps in our monitoring. NB we include the first peak for completeness and for contrast with the optically thin flares, but note that it is probably a very different sort of ejection due to its optically thick nature and accompanying X-ray behaviour. We discuss this further in Section 8.2.}
\begin{tabular}{lcccccccc}
\hline
\hline
Date & MJD & Peak-5.5GHz & Peak-9GHz & X-ray state&Rise Time &$\alpha$&$W_{min}$ &$P_{min}$\\
   (of 2010)      &          &        (mJy)        &     (mJy)     &                   &   (d)      &                 &(erg)      &   ($\mbox{erg\,s}^{-1}$)\\
\hline
Jan 21&55217.8&20.00 (0.07)&21.71 (0.04)&IS&$\le$32&0.17&$\le 2.5\times 10^{43}$&$\le 9.1 \times 10^{36}$\\
Jan 27&55223.0&18.82 (0.06)&11.87 (0.08)&SS&3&$-0.94$&1.2$\times 10^{42}$&4.5$\times 10^{36}$\\
Jan 28&55224.8&6.37 (0.43) & 4.43 (0.21) &SS&1&$-0.74 $&1.5$\times 10^{41}$&1.8$\times 10^{36}$\\
Feb 12&55239.9&16.21 (0.06)&10.83 (0.07)&SS&13&$-0.82 $&7.0$\times 10^{42}$&6.3$\times 10^{36}$\\
Feb 23&55250.8&5.93 (0.11)&4.82 (0.10)    &SS&2&$-0.42 $&3.6$\times 10^{41}$&2.1$\times 10^{36}$\\
Mar 6 & 55261.8&7.67 (0.10)&5.77 (0.07)    &SS&10&$-0.58 $&3.4$\times 10^{42}$&3.8$\times 10^{36}$\\
Mar 24 &55279.7&4.33(0.08)&3.14(0.07)      &SS&3&$-0.65 $&5.0$\times 10^{41}$&1.9$\times 10^{36}$\\
Apr 28/29 &55313/4 &1.79(0.06)&1.20(0.03)&HS&--&--&--&--\\
\hline
\end{tabular}
\label{flares}
\end{table*}

We assume that the radio source is in a state of approximate equipartition (e.g. Longair 1994) and that the minimum energy electrons are radiating at 5.5GHz, the lowest observing frequency. Then the minimum energy associated with an event is:

\begin{equation}
\frac{W_{min}}{\mbox{J}}\approx 3.0\times10^6\eta^{4/7}\bigg(\frac{V}{\mbox{m}^3} \bigg)^{3/7}\bigg(\frac{\nu}{\mbox{Hz}} \bigg)^{2/7} \bigg(\frac{L_{\nu}}{\mbox{W\,Hz}^{-1}} \bigg)^{4/7}
\end{equation}

\noindent This is equivalent to (Fender 2006):
\begin{eqnarray}
\frac{W_{min}}{\mbox{erg}}\approx3.5\times10^{33}\eta^{4/7}\bigg(\frac{\Delta t}{\mbox{s}}\bigg)^{9/7}\bigg(\frac{d}{\mbox{kpc}}\bigg)^{8/7}\bigg(\frac{\nu}{\mbox{GHz}}\bigg)^{2/7} \nonumber\\ \times\bigg(\frac{S_{\nu}}{\mbox{mJy}}\bigg)^{4/7}
\end{eqnarray}

\noindent where $\Delta t$ is the rise time, $d=3.5\mbox{kpc}$ (Shaposhnikov et al. 2010) is the distance to the source, $\nu$ is the observing frequency and $S_{\nu}$ is the flux density at frequency $\nu$. The rise times assume a spherical ejection (although we note that this may slightly over-estimate the volume compared with a continuous and approximately conical outflow) and are based on the increase of the flux density in the radio lightcurves; since the flares in the lightcurve are superimposed it is not possible to obtain exact rise times but only lower limits. We assume that the relativistic proton energies are negligible compared with those of the electrons, giving $\eta$=1. 

The minimum power is then:

\begin{equation}
\frac{P_{min}}{\mbox{erg/s}}=\bigg(\frac{W_{min}}{\mbox{erg}}\bigg)\bigg(\frac{t}{\mbox{s}}\bigg)^{-1}
\end{equation}

Resultant values of $W_{min}$ and $P_{min}$ are listed in Table~\ref{flares}. We note that this standard parametrisation for the minimum energy requires a spectral index of $\alpha = -0.75$, where $S_{\nu}=\nu^{\alpha}$; thus there is a small ($<2$) factor increase in the minimum energy to account for the variation in spectral index (see Table~\ref{flares}; Longair 1994). 

Finally, we need to account for Doppler (de-)boosting. Adopting a distance to XTE~1752$-$223 of 3.5kpc (Shaposhnikov et al. 2010), Yang et al. (2011) estimate a speed of 0.7$c$ for Component B, assuming an ejection date of MJD 55250.6. They also speculate that, if the source were situated further away, in the Galactic centre, then XTE~J1752$-$223 would be a good candidate for superluminal motion. We note that 3.5 kpc is the {\em lower} limit to the distance estimated by Ratti et al. (2012) and Chun et al. (2013 submitted) estimate a lower limit of 5kpc, based on high resolution X-ray spectroscopy and NIR photometry. Comparison with GX~339$-$4 and H1743$-322$ in Fig.~\ref{corr} implies that a distance of 8kpc is not unfeasible.

For now, following Yang et al. (2011), we adopt a relativistic value of 0.7$c$. Thus $\gamma\sim 1.4$ which gives us a range of Doppler factors $\delta=\gamma(1\mp\beta\cos \theta)^{-1}\sim0.82-4.667$, depending on whether the jet is approaching or receding and the inclination angle of the system (although we note that no receding jet was detected in the VLBI images). Thus the values of $W_{min}$ and $P_{min}$ are changed by scaling factors $\gamma\delta^{-5/7}\sim0.47-1.61$ and $\gamma\delta^{-12/7}\sim0.09-1.96$ respectively.

Considering that these calculations do not include the energy associated with bulk relativistic motion (see e.g. Fender \& Pooley (1998), Ghisellini \& Celotti (2001) for calculations which do include the motion of cold protons), include only a fairly narrow wavelength and allow for a lower distance than may be the case, these estimates can be considered conservative. It is interesting that, despite the flares apparently having a wide range of properties, they all have comparable values for minimum power. They are an order of magnitude lower than the power derived for e.g. XTE J1748$-$288 ($P\sim 8\times 10^{37}\mbox{erg\,s}^{-1}$; Brocksopp et al. 2007. See also Fender, Belloni \& Gallo (2004), and references therein, for additional comparisons between the jet power of various events), but a significant fraction of the accretion power is still required in order to launch these ejecta. We discuss in the next section whether they are all likely to be discrete ejecta.

\section{Discussion}
The outburst of XTE~J1752$-$223 was relatively conventional in some sense; it passed through the X-ray spectral states as we would expect, displayed a compact, steady jet with flat spectrum during the initial hard state period, exhibited polarised, optically thin ejection events and demonstrated some degree of correlation between the X-ray and radio emission. On the other hand, the earliest and brightest radio peak remained optically thick and presumably associated with the compact jet of the hard state, the radio emission of the decaying hard state remained optically thin, the degree of linear polarisation did not show a clear correlation with the radio behaviour and the radio flaring was not restricted to times of X-ray hardening or state transitions. Indeed, the majority of the flaring took place during the soft state when we would expect the jet source to be quenched. We address these distinctions below.

\subsection{Linking lightcurves with imaged ejecta}
\begin{figure}
\hspace*{-0.5cm}\includegraphics[width=9cm]{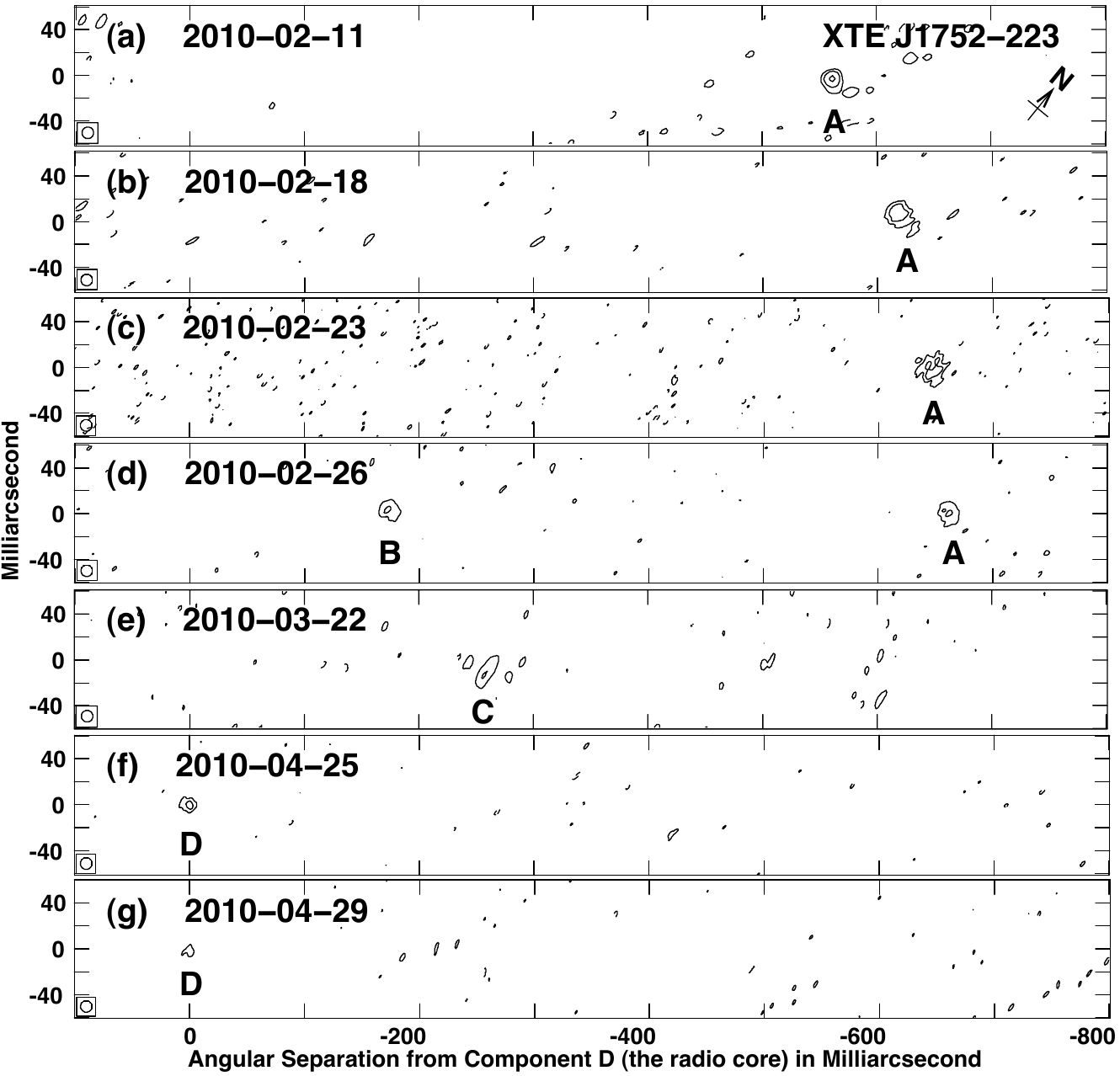}
\caption{The VLBI images of Yang et al. (2011), showing the locations of components A, B, C and d for each observing epoch.}
\label{vlbi}
\end{figure}

The series of peaks shown in Fig.~\ref{lightcurves} and Table~\ref{flares} suggests that a number of ejection events took place. The VLBI analysis of Yang et al. (2010, 2011) and Miller-Jones et al. (2011) confirm this; there are multiple components in the series of images which show proper motion and expansion. Labelled ``components A, B, C and D'' by Yang et al. (2011), where component D is the (apparently variable) core (Miller-Jones et al. 2011), we would expect the imaged ejecta to have signatures in the lightcurves also. We include the VLBI images of Yang et al. (2011) in Fig.~\ref{vlbi} for ease of comparison.

Peak 1 in the radio lightcurve is distinguished from the subsequent peaks on account of its optically thick spectrum and absence of linear polarisation. Miller-Jones et al. (2011) also discuss this peak in terms of the X-ray activity; radio flares are typically associated with the transition from a hard intermediate X-ray spectral state to a soft intermediate state (Corbel et al. 2004; Fender et al. 2004; Fender, Homan \& Belloni 2009). At the time of the first radio peak, the X-ray source had not yet softened or reduced in integrated rms variability and QPO coherence. It therefore seems likely that the first radio peak is some episode of compact jet activity occurring {\em prior} to the initial optically thin radio flare that we would expect in such an outburst (Fender et al. 2004). While the compact jets of the hard state are expected to become unstable prior to the major optically thin ejection, such instability so as to result in an optically thick peak prior to transition is uncommon, although not surprising given the increase in X-ray flux. The hard state was particularly long in this outburst, with the compact jet persisting for a couple of months, also uncommon in transient sources. It may be that this prolonged hard state created some sort of instability in the jet, leading to an optically thick flare or ejection event prior to the X-ray spectral transition. Flares associated with the compact core are not unprecedented and have been observed in SS433 (Vermeulen et al. 1993) and MAXI~J1659-152 (Paragi et al. submitted). Alternatively, and perhaps more consistent with our current understanding of transient events, it may have been the beginning of some brighter compact jet plateau phase which was then cut short by the X-ray transition to the soft state and associated quenching of the jet. Either way, such an extended jet phase probably warrants some comparison with those sources whose outbursts entirely remain in the hard state (e.g. Brocksopp, Fender, Bandyopadhay 2004; Soleri et al. 2012; Belloni et al. 2012).

Following the first radio peak, the X-ray source entered the soft state; in the standard model for black hole transient events we expect the radio emission to be quenched in this state (Corbel et al. 2004; Fender et al. 2004; Fender et al. 2009; Neilsen \& Lee 2009; Ponti et al. 2012). Any radio emission is explained in terms of additional ejecta associated with a re-crossing of the ``jet-line'' in the X-ray hardness-intensity diagram or residual emission from previous ejecta, perhaps interacting with the ISM or each other. For XTE~J1752$-$223, the sequence of radio flares which followed was both optically thin and linearly polarised. These peaks are therefore more likely than peak 1 to be associated with the ejecta imaged by the VLBI experiments. Miller-Jones et al. (2011) discuss components A and B in terms of the lightcurve peaks 2 and 4 respectively. Component A was thus thought to be ejected on MJD55218 and to decelerate via a combined model of ballistic motion followed by a Sedov phase, the ejecta decelerating rapidly as they are swept up by the ISM. 

On the other hand, Component B was more problematic; it is not detected in the images until after the fourth peak has decayed (i.e. the fourth VLBI observation), and so the association between peak 4 and component B is not compelling. Alternatively we consider the possibility that component A is actually associated with peak 4. The decay of the second peak is rapid and the ejecta had time to fade prior to the VLBI observations, thus explaining the absence of this event from each of the VLBI images. The third peak is faint and also decays quickly and so is likely to be absent from the VLBI images. However, peak 4 is bright, coincident with the VLBI images which detect component A and also decays with a similarly broad morphology to the decay of component A (Yang et al. 2011). A model of higher initial proper motion and more rapid deceleration is then required to fit the motion of component A; a shorter episode of ballistic motion, followed by the Sedov phase could still be plausible.

Consequently, component B is associated with peak 5 or 6 and component C with peak 6 or 7. Yang et al. (2011) state that component B had a higher proper motion than A, perhaps explained in terms of the detection of B being during the rise of peak 6 and so prior to any deceleration. Yang et al. (2011) also state that component C was faint, again explained by the low flux density attained by peaks 6 and 7. 

If component A was indeed associated with peak 4 then we can make an estimate of the jet speed. By the first VLBI epoch of MJD 55238.4, component A has reached an angular separation of 562.2 mas from the core. In Fig.~\ref{lightcurves}, the rise of peak 4 began around MJD 55226. This implies a proper motion of 45 mas/day, or about 1.3$c$ at 5kpc, the lower limit to the distance as estimated by Chun et al. (2013 submitted) and thus we can consider the motion to be superluminal. The initial speed would be higher still if the deceleration had already started by the time of the first epoch and/or if the distance is actually further. 

\subsection{Mechanism for flare-production}
Until this point, we have treated each optically thin peak in the lightcurve as if it were a new ejection event. This is certainly plausible, and supported by the decaying series of peaks, the still-variable core on return to the hard state, the likelihood that there is not necessarily a fixed set of X-ray properties at the onset of an ejection (Fender et al. 2009; Miller-Jones et al. 2011), the apparently discrete ejecta with different proper motions as seen in the VLBI images, and the precedence of sources such as XTE~J1859+229 and XTE~J1720$-$318 (Brocksopp et al. 2002, 2005). However, as discussed in Brocksopp et al. (2007), high levels of radio emission can also be associated with internal shocks caused by the collision of a later, faster jet and the earlier slower jet. Alternatively, shocks can form between the jet material and the surrounding environment (Corbel et al. 2002). Distinguishing between these mechanisms is not straightforward; however, a comparison between XTE~J1748$-$288 and XTE~J1752$-$223 may be warranted.

The 1998 outburst of XTE J1748$-$288 suggested that the FP was anti-correlated with the Stokes $I$ flux density during the optically thin ejection event, reaching a minimum at the outburst peak. Then the FP rose during the hard state and likely episode of collisions and shocked emission. Similarly the FP of XTE J1752$-$223 showed complex evolution during the flaring episodes, with only peaks 3, 6 and 7 having clear analogues in the FP emission. Peaks 1, 2, 4 and 5 appear to be associated with weaker polarisation. Thus we might be able to conclude that peaks 3, 6 and 7 are actually some form of interaction taking place between the jet components and/or the surrounding environment. Such a conclusion is also supported by the ongoing optically thin spectrum once the X-ray source has returned to the hard state. Assuming that component A had previously ``cleared the way'' through the surrounding medium, component C in the VLBI images could then be the emission produced in a collision between component B and previous, slower-moving material.

If it is, indeed, appropriate to compare the polarisation results of XTE J1748$-$288 and XTE J1752$-$223, bearing in mind that they both potentially reside in different environments, then we should consider the mechanism for the variability in the FP. During the initial hard state of XTE~J1752$-$223, the FP reaches a level of up to $8.0\pm1.7\%$, consistent with the maximum permitted for a self-absorbed synchrotron source ($3/(6p+13)$=10--15 \%, where $p$ is the electron energy power index; Longair 1994). Brocksopp et al. (2007) proposed that the optically thin peak of the outburst could have been dominated by a depolarised core; following the outburst peak, the increase in FP comes from the increased relative importance of a more strongly polarised jet moving away from the core. The more complex variability in XTE~J1752$-$223 is harder to explain; in addition to the core, there are multiple ejections providing different ``packets'' of differently-polarised emission, as well as additional sources of depolarisation external to the black-hole system. Any of these mechanisms, and presumably more than one, could apply in reducing the FP from its permitted maximum for optically thin emission ($(p+1)/[p+(7/3)]$=60--70\%; Longair 1994).

In a study of the 2009 outburst of H1743$-$322, Miller-Jones et al. (2012) compared VLBA images with the integrated lightcurve and found that the time of ejection coincided with the decrease in fractional rms variability and disappearance of the Type C QPO; it did not coincide with the peak of the radio flare, suggesting that there was a delay between ejection and the internal shock as the ejecta collided with the slower-moving compact jet. However, a comparison between the 2003 and 2009 outbursts showed that the X-ray timing properties at the moment of ejection are not necessarily constant, even between ejections of the same source. A similar conclusion was reached by Fender et al. (2009). It seems that the same is true of XTE J1752$-$223 -- the multiple flares show such a range of properties that it is perhaps inappropriate to assume they all arise via the same mechanism. A simple ejection might be expected to leave the core at the peak flux density; if, instead, a flare is dominated by shocked emission, we would expect the peak to occur when the ejecta is further downstream. We might even expect a broad flare, such as peak 4 to be including both. Interaction with the ISM might take place after a greater interval, depending on the distribution of gas in the environment. In a series of events, as seen in XTE~J1752$-$223, the peaks cannot all be explained in terms of the fast ejection colliding with the earlier compact jet. If, for example, component A in the VLBI images is the shocked emission between an ejection and the slower compact jet, components B and C are likely to be additional ejecta because the region of decelerated, shocked emisson is further downstream. The intervening radio flares may be the result of shocks; in a series of ejecta we would certainly expect there to be some degree of emission due to collisions.

\section{Conclusions}
The radio outburst of XTE~J1752$-$223 started with a steady, compact jet with a flat spectrum, as we would expect for a canonical black hole X-ray transient in the hard state. As the X-ray source evolved, this extended episode of jet behaviour finally gave way to a series of $\ge$7--8 peaks, the first of which remained optically thick and associated with the compact jet of the hard state. The subsequent optically thin flares showed a range of properties in terms of peak flux, duration, morphology, spectrum, polarisation, associated X-ray behaviour and energy estimate and, in attempting to determine whether the peak fluxes arise in multiple discrete ejection events, inter-ejecta collisions or ejecta-ISM collisions, we suggest that a combination of mechanisms is likely. We compare the times of flares with the ejecta imaged by VLBI facilities and try to determine which flares are associated with those imaged ejecta; this analysis suggests that at least one ejection event exhibited apparent superluminal motion. Once the source returned to the hard state, the radio emission remained optically thin, probably due to residual emission in the environment or possibly from the still-variable core. The radio source during the hard state is one of a growing list of radio-weak outliers of the standard X-ray-radio correlation for black hole X-ray transients. These observations and results highlight the need for well-sampled radio lightcurves obtained in conjunction with high-resolution images of the ejecta.

\section*{Acknowledgments}
We thanks James Miller-Jones for his help with the energetics calculations and Phil Edwards for the rapid scheduling of our many ATCA observations. The Australia Telescope is funded by the Commonwealth of Australia for operation as a National Facility managed by CSIRO. The European VLBI Network is a joint facility of European, Chinese, South African and other radio astronomy institutes funded by their national research councils. The National Radio Astronomy Observatory is a facility of the National Science Foundation operated under cooperative agreement by Associated Universities, Inc. This research has made use of data obtained through the High Energy Astrophysics Science Archive Research Center Online Service, provided by the NASA/Goddard Space Flight Center. This research has made use of MAXI data provided by RIKEN, JAXA and the MAXI team. SC  and JR have  received funding from the European Communitys Seventh Framework Programme (FP7/2007-2013) under grant agreement number ITN 215212 Black Hole Universe. SC and JR also acknowledge  support of the French Agence Nationale de la Recherche (ANR) for the CHAOS research project.

\appendix 
\section{ATCA data}

\begin{table*}
\caption{List of all ATCA data at 5.5 and 9 GHz. Errors are quoted at the 1$\sigma$ level; upper limits are quoted at the 3$\sigma$ level. Flux densities are given in mJy. Abbreviations used are HS=Hard State, IS=Intermediate State, SS=Soft State, CJ=Compact Jet, Ex=Extended emissions, E=Ejections} 
\begin{tabular}{lcccccc}
\hline
\hline
Date (UT)             &MJD    & $S$ (5.5 GHz)&$S$ (9 GHz)&Description\\

\hline
2009-10-30 & 55134.2& 2.50  $\pm$  0.08 & 2.42  $\pm$  0.05&    HS/CJ\\                         
2009-11-01 & 55136.4& 2.15  $\pm$  0.11 & 1.99  $\pm$  0.08&        HS/CJ\\                         
2009-11-05 & 55140.4& 1.87  $\pm$  0.07 & 2.05  $\pm$  0.07&               HS/CJ\\                  
2009-11-08 & 55143.3& 1.78  $\pm$  0.08 & 2.10  $\pm$  0.10&      HS/CJ\\                           
2009-12-01 & 55166.6& 1.74  $\pm$  0.14 & 1.55  $\pm$  0.07&      HS-IS?/CJ\\                           
2009-12-06 & 55171.4& 2.57  $\pm$  0.08 & 2.26  $\pm$  0.05&      HS-IS?/CJ\\                           
2009-12-18 & 55183.2& 1.88  $\pm$  0.07 & 0.84  $\pm$  0.14&      HS-IS?/CJ\\                 
2009-12-19 & 55184.9& 2.48  $\pm$  0.07 & 2.17  $\pm$  0.07&      HS-IS?/CJ\\                 
2010-01-21 & 55217.9& 20.00  $\pm$  0.06 & 21.71  $\pm$  0.04&      IS/CJ/Peak 1\\               
2010-01-24 & 55220.1& 5.59  $\pm$  0.20 & 3.30  $\pm$  0.13&          IS/CJ\\             
2010-01-24a& 55220.9& 15.52  $\pm$  0.06 & 10.76  $\pm$  0.05&      SS/E\\               
2010-01-27  &55223.0 & 18.82  $\pm$  0.06 & 11.87  $\pm$  0.08&     SS/E/Peak 2\\               
2010-01-27a& 55224.0& 5.47  $\pm$  0.12 & 2.65  $\pm$  0.14&         SS/E\\              
2010-01-28 & 55224.9& 6.37  $\pm$  0.43 & 4.43  $\pm$  0.21&        SS/E/Peak 3\\               
2010-01-30 & 55226.1& 3.17  $\pm$  0.14 & 1.45  $\pm$  0.12&        SS/E\\               
2010-02-08 & 55235.9& 12.83  $\pm$  0.04 & 8.36  $\pm$  0.07&       SS/E \\          
2010-02-12 & 55240.1& 16.21  $\pm$  0.06 & 10.83  $\pm$  0.07&      SS/E/Peak 4 \\     
2010-02-14 & 55241.9& 15.91  $\pm$  0.06 & 10.66  $\pm$  0.06&        SS/E         \\
2010-02-15 & 55242.8& 14.94  $\pm$  0.15 & 10.35  $\pm$  0.06&         SS/E        \\
2010-02-17 & 55244.8& 13.41  $\pm$  0.13 & 9.93  $\pm$  0.04&          SS/E        \\
2010-02-19 & 55246.8& 6.07  $\pm$  0.10 & 3.76  $\pm$  0.11&             SS/E      \\
2010-02-20 & 55247.9& 4.11  $\pm$  0.10 & 2.63  $\pm$  0.09&             SS/E      \\
2010-02-21 & 55248.9& 3.75  $\pm$  0.12 & 2.28  $\pm$  0.07&              SS/E     \\
2010-02-22 & 55249.9& 5.42  $\pm$  0.13 & 3.83  $\pm$  0.10&               SS/E    \\
2010-02-23 & 55250.8& 5.93  $\pm$  0.11 & 4.82  $\pm$  0.10&                   SS/E/Peak 5\\
2010-02-24 & 55251.9& 4.07  $\pm$  0.24 & 2.92  $\pm$  0.09&             SS/E      \\
2010-03-03 & 55258.9& 6.78  $\pm$  0.09 & 5.40  $\pm$  0.17&              SS/E    \\
2010-03-06 & 55261.9& 7.67  $\pm$  0.11 & 5.77  $\pm$  0.07&                   SS/E/Peak 6\\
2010-03-07 & 55262.9& 6.22  $\pm$  0.11 & 4.44  $\pm$  0.06&             SS/E      \\
2010-03-14 & 55269.8& 4.64  $\pm$  0.08 & 3.64  $\pm$  0.04&             SS/E      \\
2010-03-16 & 55271.9& 3.77  $\pm$  0.02 & 2.81  $\pm$  0.08&             SS/E      \\
2010-03-21 & 55276.8& 3.20  $\pm$  0.12 & 2.31  $\pm$  0.07&             SS/E      \\
2010-03-24 & 55279.8& 4.33  $\pm$  0.08 & 3.14  $\pm$  0.07&              SS/E/Peak 7\\
2010-03-28 & 55283.8& 4.26  $\pm$  0.08 & 3.26  $\pm$  0.04&              IS/E    \\ 
2010-03-31 & 55286.8& 3.45  $\pm$  0.04 & 2.71  $\pm$  0.04&               IS/E    \\    
2010-04-02 & 55288.9& 3.41  $\pm$  0.05 & 2.68  $\pm$  0.06&               IS/E    \\    
2010-04-03 & 55289.9& 2.99  $\pm$  0.07 & 1.91  $\pm$  0.05&                IS/E    \\   
2010-04-05 & 55291.8& 3.24 $\pm$ 0.07  & 2.44 $\pm$ 0.04&               IS/E    \\              
2010-04-06 & 55292.8& 2.81  $\pm$  0.03 & 2.01  $\pm$  0.05&                 IS/E    \\            
2010-04-14 & 55300.9& 1.07  $\pm$  0.09 & 0.90  $\pm$  0.05&                 HS/CJ+Ex?\\            
2010-04-15 & 55301.8& 1.21  $\pm$  0.03 & 0.90  $\pm$  0.03&               HS/CJ+Ex?\\               
2010-04-17 & 55303.7& 1.25  $\pm$  0.06 & 0.83  $\pm$  0.06&                   HS/CJ+Ex?\\           
2010-04-27 & 55313.7& 1.69  $\pm$  0.05 & 1.07  $\pm$  0.04&                    HS/CJ+Ex?\\          
2010-04-28 & 55314.7& 1.70  $\pm$  0.04 & 1.20  $\pm$  0.03&                HS/CJ+Ex?/Rebrightening? \\            
2010-04-29 & 55315.7& 1.79  $\pm$  0.06 & 1.13  $\pm$  0.04&                HS/CJ+Ex?/Rebrightening?\\             
2010-05-17 & 55333.9& 0.48  $\pm$  0.09 & 0.28  $\pm$  0.04&                  HS/CJ\\            
2010-05-19 & 55335.8& 0.26  $\pm$  0.02 & 0.19  $\pm$  0.04&                    HS/CJ\\          
2010-06-03 & 55350.5& 0.20  $\pm$  0.03 & 0.18  $\pm$  0.03&                   HS/CJ\\           
2010-06-12 & 55359.7& 0.19  $\pm$  0.05 & 0.15  $\pm$  0.06&                 HS/CJ\\   
2010-06-25 & 55372.5& $\le 0.06$             & $\le 0.10$            &    HS/CJ\\ 
2010-07-10 & 55387.8& $\le 0.09$             & $\le 0.08$            &    HS/CJ\\ 
\hline
\end{tabular}
\end{table*}

\begin{table*}
\caption{List of all ATCA polarisation data at 5.5 and 9 GHz. Errors are quoted at the 1$\sigma$ level. Flux densities are give in mJy, FP as a percentage and the PA in degrees.} 
\begin{tabular}{lccccccc}
\hline
\hline
Date              &MJD    &LP (5.5 GHz) &LP (9 GHz)&FP (5.5 GHz)& FP (9 GHz)&PA (5.5 GHz)&PA (9 GHz)\\

\hline
2009-10-30 & 55134.2&                      0.06 $\pm$ 0.02& 0.11 $\pm$ 0.06  &                   2.35 $\pm$ 0.99& 4.66 $\pm$ 2.58&                  12.35  $\pm$ 23.49 & 33.43  $\pm$ 56.53 \\ 		      
2009-11-01 & 55136.4&                      0.02 $\pm$ 0.03& 0.12 $\pm$ 0.04   &                  0.80 $\pm$ 1.21& 5.79 $\pm$ 1.81&                  -12.25  $\pm$ 83.62 & -27.74  $\pm$ 17.84 \\ 		      
2009-11-05 & 55140.4&                      0.18 $\pm$ 0.03& 0.13 $\pm$ 0.04    &                 9.50 $\pm$ 1.74& 6.44 $\pm$ 2.00&                  -41.70  $\pm$ 10.50 & -11.27  $\pm$ 15.90 \\ 		      
2009-11-08 & 55143.3&                      0.07 $\pm$ 0.03& 0.17 $\pm$ 0.04     &                3.87 $\pm$ 1.55& 8.05 $\pm$ 1.73&                  11.92  $\pm$ 22.81 & -31.01  $\pm$ 11.23 \\ 		      
2009-12-01 & 55166.6&                      0.02 $\pm$ 0.03& 0.02 $\pm$ 0.03   &                  1.39 $\pm$ 1.82& 1.09 $\pm$ 2.07&                  -23.45  $\pm$ 74.96 & 12.19  $\pm$ 111.33 \\ 		      
2009-12-06 & 55171.4&                      0.08 $\pm$ 0.05& 0.04 $\pm$ 0.02    &                 3.07 $\pm$ 2.04& 1.64 $\pm$ 0.68&                  -7.93  $\pm$ 20.22 & -12.57  $\pm$ 22.84 \\ 		      
2009-12-18 & 55183.2&                      0.06 $\pm$ 0.02& 0.05 $\pm$ 0.04     &                3.25 $\pm$ 1.33& 6.52 $\pm$ 4.79&                  18.51  $\pm$ 23.34 & 20.46  $\pm$ 40.93 \\ 		      
2009-12-19 & 55184.9&                      0.11 $\pm$ 0.03& 0.17 $\pm$ 0.04     &                4.27 $\pm$ 1.32& 7.60 $\pm$ 1.89&                  -3.75  $\pm$ 17.89 & 31.75  $\pm$ 15.12 \\ 		      
2010-01-21 & 55217.9&                    0.17 $\pm$ 0.03& 0.20 $\pm$ 0.04  &                   0.84 $\pm$ 0.13& 0.94 $\pm$ 0.21&                    34.12  $\pm$  9.24 & 34.46  $\pm$ 12.71 \\ 		      
2010-01-24 & 55220.1&                      0.79 $\pm$ 0.04& 0.31 $\pm$ 0.07   &                14.15 $\pm$ 0.91& 9.36 $\pm$ 2.10&                   28.50  $\pm$  3.14 & 24.39  $\pm$ 12.82 \\ 		      
2010-01-24a& 55220.9&                   0.51 $\pm$ 0.03& 0.33 $\pm$ 0.04    &                 3.31 $\pm$ 0.18& 3.09 $\pm$ 0.38&                      3.96  $\pm$  3.25 & 11.30  $\pm$  7.41 \\ 		      
2010-01-27&  355223.0 &                   0.31 $\pm$ 0.05& 0.14 $\pm$ 0.08     &                1.64 $\pm$ 0.28& 1.17 $\pm$ 0.70&                    35.37  $\pm$  9.47 & 22.22  $\pm$ 34.37 \\ 		      
2010-01-27a& 55224.0&                   0.48 $\pm$ 0.04& 0.33 $\pm$ 0.07      &               8.84 $\pm$ 0.81& 12.41 $\pm$ 2.68&                  -2.95  $\pm$  4.91 & -13.90  $\pm$ 12.59 \\ 		      
2010-01-28 & 55224.9&                     0.62 $\pm$ 0.19& 0.31 $\pm$ 0.11       &              9.67 $\pm$ 3.07& 6.92 $\pm$ 2.47&                  12.33  $\pm$ 27.72 & 36.79  $\pm$ 14.96 \\ 		      
2010-01-30 & 55226.1&                     0.09 $\pm$ 0.04& 0.26 $\pm$ 0.06        &             2.70 $\pm$ 1.16& 17.74 $\pm$ 4.52&                 19.17  $\pm$ 24.55 & -12.39  $\pm$ 14.49 \\ 		      
2010-02-08 & 55235.9&                    1.16  $\pm$0.03& 0.26 $\pm$ 0.07   &                  9.02 $\pm$ 0.24& 3.17 $\pm$ 0.84&                   -17.48  $\pm$  1.56 & 15.06  $\pm$ 14.56 \\ 		      
2010-02-12 & 55240.1&                   0.66 $\pm$ 0.03& 0.25 $\pm$ 0.04    &                 4.06 $\pm$ 0.18& 2.32 $\pm$ 0.37&                     3.98  $\pm$  2.54 & 16.87  $\pm$  9.16 \\ 		      
2010-02-14 & 55241.9&                   0.70 $\pm$ 0.04& 0.75 $\pm$ 0.07     &                4.40 $\pm$ 0.23& 7.06 $\pm$ 0.65&                    17.66  $\pm$  2.94 & 27.63  $\pm$  5.30 \\ 		      
2010-02-15 & 55242.8&                   0.34 $\pm$ 0.03& 0.23 $\pm$ 0.04      &               2.30 $\pm$ 0.17& 2.18 $\pm$ 0.34&                    -22.85  $\pm$  4.32 & -19.66  $\pm$  8.71 \\ 		      
2010-02-17 & 55244.8&                  0.45 $\pm$ 0.03& 0.27 $\pm$ 0.04 &                      3.35 $\pm$ 0.19& 2.73 $\pm$ 0.39&                   10.22  $\pm$  3.13 & 10.97  $\pm$  9.13 \\ 		      
2010-02-19 & 55246.8&                   0.43 $\pm$ 0.03& 0.24 $\pm$ 0.03  &                     7.05 $\pm$ 0.54& 6.46 $\pm$ 0.71&                  15.79  $\pm$  4.69 & 26.26  $\pm$  6.19 \\ 		      
2010-02-20 & 55247.9&                   0.33 $\pm$ 0.03& 0.27 $\pm$ 0.03   &                    8.01 $\pm$ 0.72& 10.17 $\pm$ 1.22&                 -3.50  $\pm$  3.97 & 18.67  $\pm$  6.52 \\ 		      
2010-02-21 & 55248.9&                   0.42 $\pm$ 0.03& 0.23 $\pm$ 0.03    &                 11.16 $\pm$ 0.77& 10.06 $\pm$ 1.22&                  -4.46  $\pm$  2.93 &  9.96  $\pm$  8.64 \\ 		      
2010-02-22 & 55249.9&                   0.72 $\pm$ 0.03& 0.55 $\pm$ 0.03     &                13.32 $\pm$ 0.61& 14.29 $\pm$ 0.86&                   3.94  $\pm$  1.88 & 17.88  $\pm$  3.52 \\ 		      
2010-02-23 & 55250.8&                   0.45 $\pm$ 0.03& 0.14 $\pm$ 0.04      &                 7.60 $\pm$ 0.52& 2.92 $\pm$ 0.75&                  -16.07  $\pm$  3.51 & -6.26  $\pm$ 17.47 \\ 		      
2010-02-24 & 55251.9&                   0.27 $\pm$ 0.03& 0.12 $\pm$ 0.04       &                6.53 $\pm$ 0.86& 4.16 $\pm$ 1.34&                  -18.69  $\pm$  6.66 & 35.22  $\pm$ 19.10 \\ 		      
2010-03-03 & 55258.9&                   0.97 $\pm$ 0.03& 0.73 $\pm$ 0.03        &             14.28 $\pm$ 0.51& 13.52 $\pm$ 0.72&                  29.32  $\pm$  1.82 & 36.77  $\pm$  1.83 \\ 		      
2010-03-06 & 55261.9&                   1.00 $\pm$ 0.03& 0.67 $\pm$ 0.02   &                  13.01 $\pm$ 0.44& 11.62 $\pm$ 0.41&                  38.02  $\pm$  3.62 & -42.12  $\pm$  2.01 \\ 		      
2010-03-07 & 55262.9&                   0.81 $\pm$ 0.05& 0.68 $\pm$ 0.03    &                 12.95 $\pm$ 0.76& 15.21 $\pm$ 0.62&                  34.10  $\pm$  4.52 & -43.46  $\pm$  2.01 \\ 		      
2010-03-14 & 55269.8&                   0.29 $\pm$ 0.04& 0.18 $\pm$ 0.02     &                  6.18 $\pm$ 0.92& 4.99 $\pm$ 0.51&                  -20.71  $\pm$  8.39 & -41.14  $\pm$  7.83 \\ 		      
2010-03-16 & 55271.9&                   0.46 $\pm$ 0.03& 0.10 $\pm$ 0.01   &                  12.24 $\pm$ 0.86& 3.40 $\pm$ 0.52&                   -4.11  $\pm$  2.35 & 34.01  $\pm$ 13.63 \\ 		      
2010-03-21 & 55276.8&                   0.14 $\pm$ 0.06& 0.09 $\pm$ 0.03    &                   4.38 $\pm$ 2.04& 3.91 $\pm$ 1.16&                  -13.28  $\pm$ 20.06 & -15.62  $\pm$ 15.94 \\ 		      
2010-03-24 & 55279.8&                   0.45 $\pm$ 0.03& 0.29 $\pm$ 0.02     &                10.31 $\pm$ 0.75& 9.33 $\pm$ 0.67&                   36.49  $\pm$  7.86 & 40.28  $\pm$  3.97 \\ 		      
2010-03-28 & 55283.8&                   0.34 $\pm$ 0.04& 0.15 $\pm$ 0.02      &                 7.98 $\pm$ 1.02& 4.57 $\pm$ 0.55&                  -35.13  $\pm$  9.64 & 38.78  $\pm$ 11.86 \\ 		      
2010-03-31 & 55286.8&                   0.24 $\pm$ 0.02& 0.05 $\pm$ 0.03      &                 7.00 $\pm$ 0.47& 1.80 $\pm$ 0.94&                  -43.09  $\pm$  3.71 & 28.81  $\pm$ 30.11 \\ 		      
2010-04-02 & 55288.9&                   0.09 $\pm$ 0.03& 0.11 $\pm$ 0.04       &                2.56 $\pm$ 0.83& 4.17 $\pm$ 1.57&                  -26.10  $\pm$ 18.55 & 22.63  $\pm$ 21.50 \\ 		      
2010-04-03 & 55289.9&                   0.10 $\pm$ 0.03& 0.19 $\pm$ 0.04        &               3.30 $\pm$ 1.02& 9.75 $\pm$ 2.32&                  16.29  $\pm$ 17.38 & -18.35  $\pm$ 13.56 \\ 		      
2010-04-05 & 55291.8&                         0.17 $\pm$ 0.03& 0.12 $\pm$ 0.03    &                 5.16 $\pm$ 0.79& 4.90 $\pm$ 1.29&                  -27.88  $\pm$  8.63 & -22.70  $\pm$ 15.11 \\ 		      
2010-04-06 & 55292.8&                         0.09 $\pm$ 0.02& 0.14 $\pm$ 0.03     &                3.12 $\pm$ 0.69& 6.88 $\pm$ 1.61&                  -33.50  $\pm$ 13.19 & 34.16  $\pm$ 12.81 \\ 		      
2010-04-14 & 55300.9&                         0.15 $\pm$ 0.04& 0.21 $\pm$ 0.06      &             14.30 $\pm$ 4.25& 23.72 $\pm$ 6.74&                  13.23  $\pm$ 16.77 & 18.73  $\pm$ 16.10 \\ 		      
2010-04-15 & 55301.8&                         0.07 $\pm$ 0.02& 0.05 $\pm$ 0.03       &              6.12 $\pm$ 1.47& 6.02 $\pm$ 2.80&                  -36.69  $\pm$ 13.55 & -34.35  $\pm$ 25.87 \\ 		      
2010-04-17 & 55303.7&                         0.14 $\pm$ 0.03& 0.20 $\pm$ 0.05        &           10.88 $\pm$ 2.62& 24.53 $\pm$ 6.14&                  25.84  $\pm$ 13.42 & -24.20  $\pm$ 13.88 \\ 		      
2010-04-27 & 55313.7&                         0.09 $\pm$ 0.02& 0.08 $\pm$ 0.04         &            5.41 $\pm$ 1.47& 7.51 $\pm$ 3.49&                  -21.53  $\pm$ 15.49 & 19.87  $\pm$ 26.57 \\ 		      
2010-04-28 & 55314.7&                         0.05 $\pm$ 0.02& 0.06 $\pm$ 0.03       &              3.21 $\pm$ 1.22& 5.13 $\pm$ 2.37&                  -14.57  $\pm$ 22.19 & 20.76  $\pm$ 26.44 \\ 		      
2010-04-29 & 55315.7&                         0.10 $\pm$ 0.02& 0.07 $\pm$ 0.03        &             5.35 $\pm$ 1.35& 6.33 $\pm$ 2.87&                  -34.22  $\pm$ 14.17 & 17.08  $\pm$ 25.62 \\ 		      
2010-05-17 & 55333.9&                         0.07 $\pm$ 0.04& 0.12 $\pm$ 0.04         &          13.50 $\pm$ 9.18& 41.44  $\pm$13.61&                 16.67  $\pm$ 31.25 & -6.23  $\pm$ 16.58 \\ 		      
2010-05-19 & 55335.8&                         0.11 $\pm$ 0.03& 0.09 $\pm$ 0.03          &         41.29 $\pm$ 12.40& 47.98 $\pm$ 21.11&                13.39  $\pm$ 16.12 & -23.31  $\pm$ 21.94 \\ 
2010-06-03 & 55350.5&                         0.06 $\pm$ 0.02& 0.07 $\pm$ 0.02           &        29.33 $\pm$ 10.29& 40.04 $\pm$ 15.57&                -25.36  $\pm$ 18.66 & -0.15  $\pm$   20.17 \\ 
\hline            
\end{tabular}     
\end{table*}


\begin{thebibliography}{}
\bibitem[\protect\citeauthoryear{Belloni}{2010}]{} Belloni T. M., 2010, in Belloni T., ed., Lecture Notes in Physics, Vol. 794, The Jet Paradigm -- From Microquasars to Quasars, p.53
\bibitem[\protect\citeauthoryear{Belloni}{2012}]{} Belloni T.M. et al., 2012, ATel. 4450
\bibitem[\protect\citeauthoryear{Brocksopp et al.}{2005}]{2005MNRAS.356..125B} Brocksopp C., Corbel S., Fender R.~P., Rupen M., Sault R., Tingay S.~J., Hannikainen D., O'Brien K., 2005, MNRAS, 356, 125 
\bibitem[\protect\citeauthoryear{Brocksopp et al.}{2009}]{2009ATel.2278....1B} Brocksopp C., Corbel S., Tzioumis T., Fender R., 2009, ATel, 2278
\bibitem[\protect\citeauthoryear{Brocksopp et al.}{2010}]{2010ATel.2400....1B} Brocksopp C., Corbel S., Tzioumis T., Fender R., Coriat M., 2010a, ATel, 2400
\bibitem[\protect\citeauthoryear{Brocksopp, Bandyopadhyay, \& Fender}{2004}]{2004NewA....9..249B} Brocksopp C., Bandyopadhyay R.~M., Fender R.~P., 2004, NewA, 9, 249
\bibitem[\protect\citeauthoryear{Brocksopp et al.}{2007}]{2007MNRAS.378.1111B} Brocksopp C., Miller-Jones J.~C.~A., Fender R.~P., Stappers B.~W., 2007, MNRAS, 378, 1111 
\bibitem[\protect\citeauthoryear{Brocksopp et al.}{2010}]{2010ATel.2438....1B} Brocksopp C., Yang J., Corbel S., Tzioumis T., Fender R., 2010b, ATel, 2438
\bibitem[\protect\citeauthoryear{Brocksopp et al.}{2010}]{2010ATel.2439....1B} Brocksopp C., Yang J., Corbel S., Tzioumis T., Fender R., 2010c, ATel, 2439
\bibitem[\protect\citeauthoryear{Brocksopp et al.}{2002}]{2002MNRAS.331..765B} Brocksopp C., et al., 2002, MNRAS, 331, 765 
\bibitem[\protect\citeauthoryear{Buxton \& Bailyn}{2004}]{2004ApJ...615..880B} Buxton M.~M., Bailyn C.~D., 2004, ApJ, 615, 880 
\bibitem[\protect\citeauthoryear{Buxton et al.}{2010}]{2010ATel.2549....1B} Buxton M., Dincer T., Kalemci E., Tomsick J., 2010, ATel, 2549
\bibitem[\protect\citeauthoryear{Cadolle Bel}{2007}]{} Cadolle Bel M. et al., ApJ, 659, 549 
\bibitem[\protect\citeauthoryear{Corbel et al.}{2000}]{} Corbel S., Fender R.~P., Tzioumis A.~K., Nowak M., McIntyre V., Durouchoux P., Sood R., 2000, A\&A, 359, 251 
\bibitem[\protect\citeauthoryear{Corbel et al.}{2002}]{2002Sci...298..196C} Corbel S., Fender R.~P., Tzioumis A.~K., Tomsick J.~A., Orosz J.~A., Miller J.~M., Wijnands R., Kaaret P., 2002, Sci, 298, 196 
\bibitem[\protect\citeauthoryear{Corbel et al.}{2003}]{} Corbel S., Nowak M.~A., Fender R.~P., Tzioumis A.~K., Markoff S., 2003, A\&A, 400, 1007 
\bibitem[\protect\citeauthoryear{Corbel et al.}{2004}]{2004ApJ...617.1272C} Corbel S., Fender R.~P., Tomsick J.~A., Tzioumis A.~K., Tingay S., 2004, ApJ, 617, 1272
\bibitem[\protect\citeauthoryear{Corbel et al.}{2013a}]{2013a} Corbel S., Coriat M., Brocksopp C., Tzioumis T., Fender R.P., Tomsick J.A., Buxton M., Bailyn C., 2013a, MNRAS, 428, 2500
\bibitem[\protect\citeauthoryear{Corbel et al.}{2013b}]{2013b} Corbel S. et al., MNRAS submitted
\bibitem[\protect\citeauthoryear{Coriat et al.}{2009}]{2009MNRAS.400..123C} Coriat M., Corbel S., Buxton M.~M., Bailyn C.~D., Tomsick J.~A., K{\"o}rding E., Kalemci E., 2009, MNRAS, 400, 123
\bibitem[\protect\citeauthoryear{Coriat et al.}{2011}]{2011MNRAS.414..677C} Coriat M., et al., 2011, MNRAS, 414, 677 
\bibitem[\protect\citeauthoryear{Corral-Santana, Casares, \& Rodr{\'{\i}}guez-Gil}{2010}]{2010ATel.2804....1C} Corral-Santana J.~M., Casares J., Rodr{\'{\i}}guez-Gil P., 2010a, ATel, 2804
\bibitem[\protect\citeauthoryear{Corral-Santana, Casares, \& Rodr{\'{\i}}guez-Gil}{2010}]{2010ATel.2805....1C} Corral-Santana J.~M., Casares J., Rodr{\'{\i}}guez-Gil P., 2010b, ATel, 2805
\bibitem[\protect\citeauthoryear{Corral-Santana et al.}{2010}]{2010ATel.2818....1C} Corral-Santana J.~M., Rodr{\'{\i}}guez-Gil P., Guerra J.~C., Casares J., 2010c, ATel, 2818 
\bibitem[\protect\citeauthoryear{Curran et al.}{2010}]{2010ATel.2424....1C} Curran P.~A., Evans P.~A., Still M., Brocksopp C., Done C., 2010, ATel, 2424
\bibitem[\protect\citeauthoryear{Curran et al.}{2011}]{2011MNRAS.410..541C} Curran P.~A., Maccarone T.~J., Casella P., Evans P.~A., Landsman W., Krimm H.~A., Brocksopp C., Still M., 2011, MNRAS, 410, 541 
\bibitem[\protect\citeauthoryear{Fender \& Pooley}{1998}]{1998MNRAS.300..573F} Fender R.~P., Pooley G.~G., 1998, MNRAS, 300, 573
\bibitem[\protect\citeauthoryear{Fender}{2001}]{2001MNRAS.322...31F} Fender R.~P., 2001, MNRAS, 322, 31
\bibitem[\protect\citeauthoryear{Fender}{2003}]{} Fender R. P., 2003, Ap\&SS, 288, 79
\bibitem[\protect\citeauthoryear{Fender, Belloni, \& Gallo}{2004}]{2004MNRAS.355.1105F} Fender R.~P., Belloni T.~M., Gallo E., 2004, MNRAS, 355, 1105 
\bibitem[\protect\citeauthoryear{Fender}{2006}]{} Fender R. P., 2006, in Walter L., and van der Klis M., ed., Compact Stellar X-ray Sources. Cambridge Univ. Press, Cambridge
\bibitem[\protect\citeauthoryear{Fender et al.}{1999}]{1999MNRAS.304..865F} Fender R.~P., Garrington S.~T., McKay D.~J., Muxlow T.~W.~B., Pooley G.~G., Spencer R.~E., Stirling A.~M., Waltman E.~B., 1999, MNRAS, 304, 865 
\bibitem[\protect\citeauthoryear{Fender, Homan, \& Belloni}{2009}]{2009MNRAS.396.1370F} Fender R.~P., Homan J., Belloni T.~M., 2009, MNRAS, 396, 1370 
\bibitem[\protect\citeauthoryear{Gallo, Fender, \& Pooley}{2003}]{2003MNRAS.344...60G} Gallo E., Fender R.~P., Pooley G.~G., 2003, MNRAS, 344, 60 
\bibitem[\protect\citeauthoryear{Gallo, Miller, \& Fender}{2012}]{2012MNRAS.423..590G} Gallo E., Miller B.~P., Fender R., 2012, MNRAS, 423, 590
\bibitem[\protect\citeauthoryear{Ghisellini \& Celotti}{2001}]{2001MNRAS.327..739G} Ghisellini G., Celotti A., 2001, MNRAS, 327, 739
\bibitem[\protect\citeauthoryear{Hannikainen et al.}{1998}]{1998A&A...337..460H} Hannikainen D.~C., Hunstead R.~W., Campbell-Wilson D., Sood R.~K., 1998, A\&A, 337, 460 
\bibitem[\protect\citeauthoryear{Homan}{2010}]{2010ATel.2387....1H} Homan J., 2010, ATel, 2387
\bibitem[\protect\citeauthoryear{Laurent et al.}{2011}]{2011Sci...332..438L} Laurent P., Rodriguez J., Wilms J., Cadolle Bel M., Pottschmidt K., Grinberg V., 2011, Sci, 332, 438 
\bibitem[\protect\citeauthoryear{Longair}{1994}]{}Longair M. S., 1994, High Energy Astrophysics, Vol. 2. Cambridge Univ. Press, Cambridge
\bibitem[\protect\citeauthoryear{Markwardt et al.}{2009}]{2009ATel.2258....1M} Markwardt C.~B., et al., 2009a, ATel, 2258
\bibitem[\protect\citeauthoryear{Markwardt et al.}{2009}]{2009ATel.2261....1M} Markwardt C.~B., Barthelmy S.~D., Evans P.~A., Swank J.~H., 2009b, ATel, 2261
\bibitem[\protect\citeauthoryear{Miller-Jones et al.}{2011}]{2011MNRAS.415..306M} Miller-Jones J.~C.~A., Jonker P.~G., Ratti E.~M., Torres M.~A.~P., Brocksopp C., Yang J., Morrell N.~I., 2011, MNRAS, 415, 306 
\bibitem[\protect\citeauthoryear{Miller-Jones et al.}{2012}]{2012MNRAS.421..468M} Miller-Jones J.~C.~A., et al., 2012, MNRAS, 421, 468
\bibitem[\protect\citeauthoryear{Mu{\~n}oz Darias et al.}{2010}]{2010ATel.2518....1M} Mu{\~n}oz Darias T., Motta S., Belloni T., Homan J., 2010b, ATel, 2518
\bibitem[\protect\citeauthoryear{Mu{\~n}oz-Darias et al.}{2010}]{2010MNRAS.404L..94M} Mu{\~n}oz-Darias T., Motta S., Pawar D., Belloni T.~M., Campana S., Bhattacharya D., 2010a, MNRAS, 404, L94 
\bibitem[\protect\citeauthoryear{Nakahira et al.}{2009}]{2009ATel.2259....1N} Nakahira S., et al., 2009, ATel, 2259
\bibitem[\protect\citeauthoryear{Nakahira et al.}{2010}]{2010PASJ...62L..27N} Nakahira S., et al., 2010, PASJ, 62, L27 
\bibitem[\protect\citeauthoryear{Nakahira et al.}{2012}]{2012PASJ...64...13N} Nakahira S., et al., 2012, PASJ, 64, 13 
\bibitem[\protect\citeauthoryear{Neilsen \& Lee}{2009}]{2009Natur.458..481N} Neilsen J., Lee J.~C., 2009, Natur, 458, 481 
\bibitem[\protect\citeauthoryear{Ponti et al.}{2012}]{2012MNRAS.422L..11P} Ponti G., Fender R.~P., Begelman M.~C., Dunn R.~J.~H., Neilsen J., Coriat M., 2012, MNRAS, 422, L11
\bibitem[\protect\citeauthoryear{Ratti et al.}{2012}]{2012MNRAS.tmp.3053R} Ratti E.~M., et al., 2012, MNRAS, 3053
\bibitem[\protect\citeauthoryear{Reis et al.}{2011}]{2011MNRAS.410.2497R} Reis R.~C., et al., 2011, MNRAS, 410, 2497 
\bibitem[\protect\citeauthoryear{Remillard \& The ASM Team at MIT}{2009}]{2009ATel.2265....1R} Remillard R.~A. et al., 2009, ATel, 2265
\bibitem[\protect\citeauthoryear{Rodriguez et al.}{2008}]{2008ApJ...675.1449R} Rodriguez J., et al., 2008a, ApJ, 675, 1449 
\bibitem[\protect\citeauthoryear{Rodriguez et al.}{2008}]{2008ApJ...675.1436R} Rodriguez J., et al., 2008b, ApJ, 675, 1436 
\bibitem[\protect\citeauthoryear{Russell et al.}{2007}]{2007MNRAS.379.1401R} Russell D.~M., Maccarone T.~J., K{\"o}rding E.~G., Homan J., 2007, MNRAS, 379, 1401 
\bibitem[\protect\citeauthoryear{Russell et al.}{2010}]{2010ATel.2775....1R} Russell D.~M., Mu{\~n}oz-Darias T., Lewis F., Soleri P., 2010, ATel, 2775
\bibitem[\protect\citeauthoryear{Russell et al.}{2012}]{2012MNRAS.419.1740R} Russell D.~M., et al., 2012, MNRAS, 419, 1740 
\bibitem[\protect\citeauthoryear{Shaposhnikov, Markwardt, \& Swank}{2009}]{2009ATel.2269....1S} Shaposhnikov N., Markwardt C.~B., Swank J.~H., 2009, ATel, 2269
\bibitem[\protect\citeauthoryear{Shaposhnikov}{2010}]{2010ATel.2391....1S} Shaposhnikov N., 2010a, ATel, 2391
\bibitem[\protect\citeauthoryear{Shaposhnikov et al.}{2010}]{2010ApJ...723.1817S} Shaposhnikov N., Markwardt C., Swank J., Krimm H., 2010b, ApJ, 723, 1817
\bibitem[\protect\citeauthoryear{Soleri et al.}{2012}]{}Soleri P. et al., 2012, MNRAS in press  
\bibitem[\protect\citeauthoryear{Torres et al.}{2009}]{2009ATel.2263....1T} Torres M.~A.~P., Jonker P.~G., Steeghs D., Yan H., Huang J., Soderberg A.~M., 2009a, ATel, 2263
\bibitem[\protect\citeauthoryear{Torres et al.}{2009}]{2009ATel.2268....1T} Torres M.~A.~P., Steeghs D., Jonker P.~G., Thompson I., Soderberg A.~M., 2009b, ATel, 2268
\bibitem[\protect\citeauthoryear{Vermeulen et al.}{1993}]{1993A&A...270..177V} Vermeulen R.~C., Schilizzi R.~T., Spencer R.~E., Romney J.~D., Fejes I., 1993, A\&A, 270, 177
\bibitem[\protect\citeauthoryear{Wilson-Hodge et al.}{2009}]{2009ATel.2280....1W} Wilson-Hodge C.~A., Camero-Arranz A., Case G., Chaplin V., Connaughton V., 2009, ATel, 2280
\bibitem[\protect\citeauthoryear{Yang et al.}{2010}]{2010MNRAS.409L..64Y} Yang J., Brocksopp C., Corbel S., Paragi Z., Tzioumis T., Fender R.~P., 2010, MNRAS, 409, L64 
\bibitem[\protect\citeauthoryear{Yang et al.}{2011}]{2011MNRAS.418L..25Y} Yang J., Paragi Z., Corbel S., Gurvits L.~I., Campbell R.~M., Brocksopp C., 2011, MNRAS, 418, L25 

\end{thebibliography}
\end{document}